\documentclass[final,3p,times,authoryear, sort&compress]{elsarticle}

\usepackage{lineno,hyperref}
\hypersetup{colorlinks=true, linkcolor=blue,urlcolor=blue,citecolor=blue}
\usepackage{soul}
\usepackage{color, xcolor}
\usepackage{amsmath}
\usepackage{amssymb}
\usepackage{MnSymbol}
\usepackage{graphicx}
\usepackage[ruled,linesnumbered]{algorithm2e}
\usepackage{algpseudocode}
\usepackage{epstopdf}
\usepackage{float}
\usepackage{array}
\usepackage{booktabs}
\usepackage{tabu}
\usepackage{longtable}
\usepackage{subfig}
\usepackage{multirow} 
\usepackage{makecell}
\usepackage{bbding} 

\usepackage{natbib}

\makeatletter
\def\algbackskip{\hskip-\ALG@thistlm}
\makeatother

\sethlcolor{yellow}
\soulregister{\cite}7 
\soulregister{\citep}7
\soulregister{\citet}7 
\soulregister{\ref}7 
\soulregister{\pageref}7 

\journal{Elsevier}

\biboptions{authoryear}

\begin{document}
	
	\begin{frontmatter}
		
		\title{A Transparent and Nonlinear Method for Variable Selection \tnoteref{mytitlenote}}
		
		\author[add1,add2]{Keyao Wang}
		\ead{kywang@buaa.edu.cn}
		\author[add1,add3]{Huiwen Wang}
		\ead{wanghw@via.sina.com}
		\author[add1]{Jichang Zhao\corref{mycorrespondingauthor}}
		\ead{jichang@buaa.edu.cn}
		\author[add4]{Lihong Wang}
		\ead{wlh@isc.org.cn}
		
		\address[add1]{School of Economics and Management, Beihang University, Beijing, China}
		\address[add2]{Beijing Key Laboratory of Emergency Support Simulation Technologies of City Operations, Beijing, China}
		\address[add3]{Key Laboratory of Complex System Analysis, Management and Decision (Beihang University), Ministry of Education, China}
		\address[add4]{National Computer Network Emergency Response Technical Team/Coordination Center of China, Beijing, China}
		
		\cortext[mycorrespondingauthor]{Corresponding author.}
		
		\begin{abstract}
			Variable selection is a procedure to attain the truly important predictors from inputs. Complex nonlinear dependencies and strong coupling pose great challenges for variable selection in high-dimensional data. In addition, real-world applications have increased demands for interpretability of the selection process. A pragmatic approach should not only attain the most predictive covariates, but also provide ample and easy-to-understand grounds for removing certain covariates. In view of these requirements, this paper puts forward an approach for transparent and nonlinear variable selection. In order to transparently decouple information within the input predictors, a three-step heuristic search is designed, via which the input predictors are grouped into four subsets: the relevant to be selected, and the uninformative, redundant, and conditionally independent to be removed. A nonlinear partial correlation coefficient is introduced to better identify the predictors which have nonlinear functional dependence with the response. The proposed method is model-free and the selected subset can be competent input for commonly used predictive models. Experiments demonstrate the superior performance of the proposed method against the state-of-the-art baselines in terms of prediction accuracy and model interpretability.

		\end{abstract}
	
		\begin{keyword}
			Variable selection \sep High-dimensional \sep Interpretation \sep Nonlinear relevance 
		\end{keyword}
		
	\end{frontmatter}
	
	\newpageafter{abstract}

	\section{Introduction} \label{introduction}
	Predictive modeling often encounters high-dimensional data \citep{yinAdaptiveFeatureSelection2022, hossnyFeatureSelectionMethods2020, chaudhariNeuralNetworkSystems2023, luWhatMattersfor2023}, and selecting the truly important predictors is the key to achieving accurate and reliable predictions \citep{guyonIntroductionVariableFeature2003}. However, variable selection often encounters great challenges from complex nonlinearity and strong coupling that are widespread in high-dimensional data \citep{hastieBasisExpansionsRegularization2001}. Variable selection methods, e.g., feature screening \citep{fanSureIndependenceScreening2008, liFeatureScreeningDistance2012}, and stepwise \citep{efroymsonMultipleRegressionAnalysis1960, buhlmannVariableSelectionHighdimensional2010}, have been devoted to selecting the subset of predictors which the response is most related to, with the expectation of improving prediction accuracy, and reducing computational cost.  Although improving the transparency of variable selection has important implications in enhancing model interpretability \citep{murdochDefinitionsMethodsApplications2019, rudinInterpretableMachineLearning2022}, it has seldomly mentioned in high-dimensional predictions.
	
	\textit{How to transparently decouple the information contained in the inputs} is an important, yet easily overlooked area of concern when designing variable selection methods.
	Collected based on limited experience, the input set of predictors of high-dimensional data are usually ``dirty" \citep{caiFeatureSelectionMachine2018}. Only several predictors may be truly relevant to predict the response, and the rest may be lack of information, convey the same information as other predictors or carry irrelevant information to the response \citep{wanR2CIInformationTheoreticguided2022}.
	To achieve transparent information decoupling, two issues need further discussion. \textit{(i)} how to effectively select the truly important predictors; \textit{(ii)} how to transparently delete predictors which are of no use for the prediction.
	
	When selecting the predictors which are useful for predicting the response, most prevailing approaches for variable selection usually assume that the response and predictors follow some simple forms of functional dependencies, e.g., linear \citep{tibshiraniRegressionShrinkageSelection1996}, monotonic \citep{zhuModelFreeFeatureScreening2011}, or additive \citep{marraPracticalVariableSelection2011}. In reality, however, there are more diverse and complicated nonlinear forms of dependencies, e.g., nonmonotonic or even oscillatory functional dependencies between predictors and response \citep{chatterjeeNewCoefficientCorrelation2021}, and the interactions among predictors \citep{wanDynamicInteractionFeature2021}. Traditional approaches have difficulty in effectively identifying and selecting such complex nonlinear dependencies. The omission of some key nonlinear relevant predictors can greatly damage the accuracy of predictive modeling \citep{azadkiaSimpleMeasureConditional2021a}.
	
	When deleting the predictors which are of no use to predict the response, prevailing approaches mainly rank the predictors based on their correlations to the response, and divide the predictors into two categories, i.e., those relevant to the response, and those independent of the response \citep{songFeatureSelectionBased2017, dessiSimilarityFeatureSelection2015}. Most approaches make assumptions that the inputs exclude predictors without information, or there is no multicollinearity between predictors \citep{fanStatisticalFoundationsData2020}. However, the uninformative predictors and collinearity widely exist in real-world applications \citep{liFeatureSelectionData2017}. It is difficult for classical approaches to categorize and remove multiple types of predictors separately, which leads to intransparent selection process, and diminishes their efficiency and interpretability.
	
	To alleviate these challenges in selecting and deleting predictors, this article constructs a Transparent and Nonlinear Variable Selection (TNVS) method for high-dimensional data. The input predictors are divided into four nonoverlapping subsets to achieve Transparent Information Decoupling (TID), i.e., the relevant predictors to be selected, and the uninformative, redundant and conditionally independent predictors to be deleted. The transparent selection and deletion improve predictive accuracy and model interpretability. The main contributions are as follows.
	\begin{enumerate}[(1)]
		\item Equipped with the recently proposed nonlinear partial correlation, TNVS is able to select predictors with a diversity of complex nonlinear relevance to the response, including nonmonotonic or oscillatory functional dependence between the response and predictors, and interactions among predictors.
		\item Information entropy is adopted to filter out uninformative predictor, Gram-Schmidt orthogonalization is adopted to remove redundant predictors which are collinear with the relevant predictors, and the nonlinear partial correlation coefficient is adopted to remove the conditionally independent predictors. In this way, the predictors are classified into different types, and the reasons for removing certain predictors are clearly indicated.
		\item The effectiveness and interpretability of the proposed method are demonstrated against the state-of-the-art baselines. The proposed method categorizes the predictors with high accuracy on high-dimensional nonlinear simulations, and the selected subset of predictors improve out-of-sample predictions on real datasets.
	\end{enumerate}
	
	The remainder of the paper is organized as follows. Section \ref{Literature} reviews some prevailing methods in related realms. Section \ref{Preliminary} describes the concepts involved in information decoupling and the proxy measures. Section \ref{method} presents the search framework of the proposed TNVS. Section \ref{Simulations} demonstrates the effectiveness and interpretability of the proposed method on simulation problems. The performance of the proposed method on real data applications is discussed in Section \ref{Realdata}, including the predictive effectiveness of the selected subset, model interpretability, post hoc interpretability, and tuning parameter stability. Conclusions and future work are summarized in Section \ref{Conclusion}.

	\section{Related work}\label{Literature}
	Since our purpose is to derive a transparent variable selection method for high-dimensional data with nonlinear functional dependencies, we reviewed the most relevant methods in this section, including feature screening approaches based on correlations, feature screening approaches based on partial correlations, and stepwise selection methods based on partial correlations.
	
	\textbf{Correlation coefficients}, and other statistical measures for dependencies between two variables, have been widely introduced into feature screening to identify the nonlinear relevance between the response and the predictor in high-dimensional data. Some feature screening methods tried to adopt nonlinear correlations, e.g., generalized correlation \citep{hallUsingGeneralizedCorrelation2009}, distance correlation \citep{liFeatureScreeningDistance2012}, ball correlation \citep{panGenericSureIndependence2019}, and projection correlation \citep{liuModelFreeFeatureScreening2022}. Others attempted to enhance the Pearson’s correlation \cite{zhuModelFreeFeatureScreening2011} and the leverage score of singular value decompositions \citep{zhongModelfreeVariableScreening2021} through the slicing scheme and inverse regression idea \citep{liSlicedInverseRegression1991}. However, it is difficult for these methods to recruit those predictors that have weak marginal utilities but are jointly strongly correlated with the response. In this study, we turn to adopt the partial correlation rather than the correlations to better detect the interacted covariates.
	
	\textbf{Feature screening methods based on Partial correlations (PC)} were able to avoid the false omission of some important predictors \citep{barutConditionalSureIndependence2016, wangSureIndependenceScreening2017}. Given a set of known key variables, these methods could detect and select the important predictors on which the response was strongly dependent through a one-pass screening and ordering algorithm. Meanwhile, the predictors which the response was conditionally independent of were screened out. However, given the newly enlarged selected predictor set, some removed predictors were likely to become jointly correlated with the response. This indicates that if complex interactions exist among the predictors, the selection may still be insufficient to capture all the important predictors. In addition, such methods require prior knowledge of key variables, but such knowledge is usually unobtainable in practical problems. This stimulates the design of stepwise methods for more sufficient selection and more generalized initialization.
	
	\textbf{Stepwise variable selection based on PCs} can be a substitute when the predictors have complex interactions or there is little prior knowledge of the key variables, e.g., the PC-simple algorithm \citep{buhlmannVariableSelectionHighdimensional2010} and the thresholded partial correlation (TPC) \citep{liVariableSelectionPartial2017}. Nevertheless, a limitation of PC-simple and TPC is that they are both designed with the linear partial correlation coefficient, and thus they are solid only in linear regression models. This motivates us to introduce nonlinear PCs into stepwise methods, so that these methods can better identify the nonlinear conditional relevance between the response and predictors, and be extended to more general settings.
	
	\textbf{A nonlinear partial correlation coefficient}, namely Conditional Dependence Coefficient (CODEC) \citep{azadkiaSimpleMeasureConditional2021a}, was a recent significant development in this field. CODEC was a fully nonparametric measure of the nonlinear conditional dependence between two variables given a set of other variables. Based on the CODEC, a forward selection method was further presented, namely Feature Ordering by Conditional Independence (FOCI). FOCI could identify the conditionally dependent predictors from a diversity of complicated associations. However, the unimportant predictors were evaluated in each iteration and not removed until the end of the search, which is time-consuming. Besides, the removed subset was considered as a whole, though it may include various types of predictors, which is hard to interpret. 
	Our rationale is that by introducing multiple time-saving measures to remove the unimportant predictors during the search, the subset to be evaluated will shrink much faster than FOCI, which increases efficiency. Moreover, the removed subset can be divided in a quasi-chunked fashion, which increases interpretability. 
	
	The related methods are summarized in Table \ref{tab:related_work}. Prevailing feature screening methods based on correlations are listed, including Sure Independent Screening (SIS), Sure Independent Ranking and Screening (SIRS), Distance Correlation based Sure Independence Screening (DC-SIS), and Weighted Leverage Score (WLS). Stepwise methods based on partial correlations are also considered, including PC-simple, TPC, and FOCI. From Table \ref{tab:related_work}, it is observed that a nonlinear method which can both select nonmonotonic nonlinear relevance, and remove the uninformative and redundant predictors, does not exist. A fully interpretable method, which can transparently select and delete certain types of predictors, is highly needed. These motivate us to present a transparent manner to decouple complex information for high-dimensional data, and design an effective and interpretable scheme for variable selection. To our knowledge, ours is the first report to design variable selection method directed by transparent information decoupling.

	\begin{table}[H]
		\centering
		\footnotesize
		\caption{A selective list of the variable selection approaches related to the proposed method.}
		\begin{tabular}{lccccc}
			\toprule
			Category & Method & Nonlinear relevant & Redundant & Uninformative & Interpretable \\
			\midrule
			\multirow{4}{*}{\makecell[l]{Correlation, \\ feature screening}}  & SIS \citep{fanSureIndependenceScreening2008} &  ×   & ×     & $\surd$ & Partially \\
			& SIRS \citep{zhuModelFreeFeatureScreening2011} &  Monotonic  & ×     & $\surd$ & Partially \\
			& DC-SIS \citep{liFeatureScreeningDistance2012} & Monotonic  & ×     & $\surd$ & Partially \\
			& WLS \citep{zhongModelfreeVariableScreening2021} & $\surd$    & ×     & $\surd$ & Partially \\
			\midrule
			\multirow{3}{*}{\makecell[l]{Partial correlation, \\ stepwise}}  & PC-simple \citep{buhlmannVariableSelectionHighdimensional2010} & ×  & $\surd$   & × & Partially \\
			& TPC \citep{liVariableSelectionPartial2017} & ×   & $\surd$     & × & Partially \\
			& FOCI \citep{azadkiaSimpleMeasureConditional2021a} & $\surd$  & $\surd$ (inefficient)    & × & Partially \\
			& proposed TNVS & $\surd$  & $\surd$ & $\surd$ & Fully \\
			\bottomrule
		\end{tabular}%
		\label{tab:related_work}%
	\end{table}%

	\section{Information decoupling and the proxy measures} \label{Preliminary}
	
\begin{figure}[htbp]
		\centering
		\includegraphics[width=0.75\linewidth]{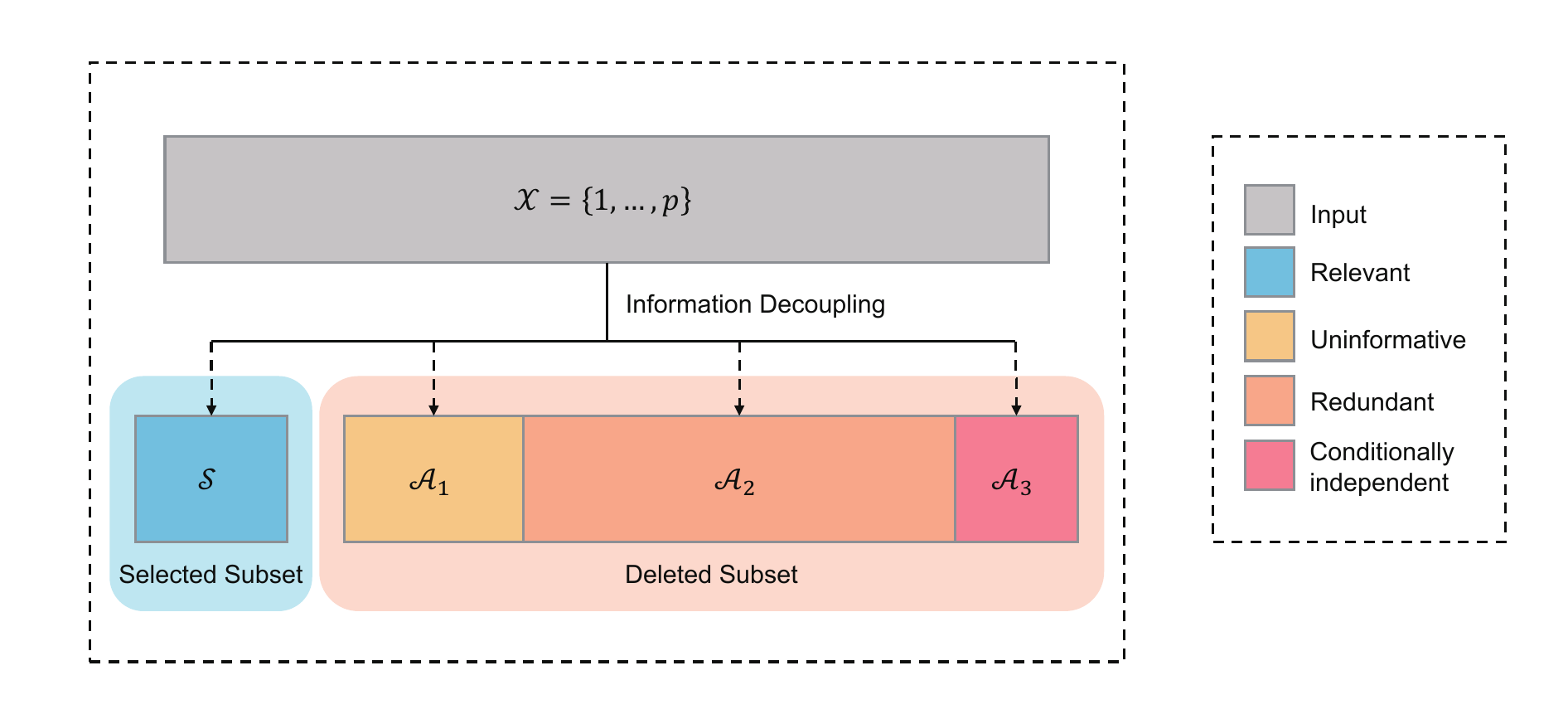}
		\caption{The input predictors are transparently grouped into four disjoint parts. The index set of the inputs is $\mathcal{X}=\{1,~2,\cdots,p\}$, which is divided into the relevant subset $\mathcal{S}$, the uninformative subset $\mathcal{A}_1$, the redundant subset $\mathcal{A}_2$, and the conditionally independent subset $\mathcal{A}_3$.}
		\label{fig:aim}
	\end{figure}

	The proposed variable selection can be regarded as an information decoupling process of the input set of predictors. The input set is transparently divided into four disjoint subsets, i.e., the subset $\mathcal{S}$ which is relevant to the response, the uninformative subset $\mathcal{A}_1$, the redundant subset $\mathcal{A}_2$, and the conditionally independent subset $\mathcal{A}_3$, as shown in Fig. \ref{fig:aim}. In this section, the four subsets in information decoupling are defined, and their corresponding measures in nonlinear supervised learning are described. An example of the four types of predictors are given in Appendix \ref{subsec:A1}.
	
		\subsection{The relevant subset and its measure}
	Let $\mathbf{Y}=(y_1,\cdots,y_n)^{\prime}$ be the response, and $\mathbf{X}=(\mathbf{X}_1,\cdots,\mathbf{X}_p)$ be the $p$-dimensional input predictors, where $\mathbf{X}_j=(x_{1j},\cdots,x_{nj})^{\prime}$, $j=1,\cdots,p$, and $n$ is the sample size. Let $\mathcal{X}=\{1,\ 2,\cdots,p\}$ be the index set of the input predictors. For any subset of indices $\mathcal{S}\subseteq\mathcal{X}$, let $\mathbf{X}_{\mathcal{S}}$ be the data matrix composed of all the $\mathbf{X}_j$ that satisfies $j\in\mathcal{S}$. For an $\mathcal{S}$, if there is a function $f(\cdot)$ that makes $\mathbf{Y}=f(\mathbf{X}_{\mathcal{S}})+\mathbf{\varepsilon}$, where $\mathbf{\varepsilon}$ is a stochastic error, then $\mathcal{S}$ stands for the index set of the true model, which is named the relevant subset in this paper. In high-dimensional data, $p$ is close to or even larger than $n$. In such cases, sparsity assumption usually holds that the dimension of $\mathbf{X}_\mathcal{S}$ is much less than $p$.

	In real problems, $\mathbf{Y}$ and $\mathbf{X}_\mathcal{S}$ can be linearly correlated, or they can also be associated in the form of a variety of complex nonlinear functions. To date, handling complex nonlinear correlations remains a great challenge \citep{fanStatisticalFoundationsData2020}. The primary goal of the proposed variable selection method is to obtain the correlated subset $\mathcal{S}$ from the input set $\mathcal{X}$. For a given group of predictors $\mathbf{X}_\mathcal{G}$, $ \mathcal{G}\subseteq\mathcal{X}$, and an unknown predictor $\mathbf{X}_j$, $j\in \mathcal{X}\ \backslash\ \mathcal{G}$, the Conditional Dependence Coefficient (CODEC) \citep{azadkiaSimpleMeasureConditional2021a} can measure the nonlinear conditional correlation between $\mathbf{Y}$ and $\mathbf{X}_j$ given $\mathbf{X}_\mathcal{G}$, as well as the interaction between $\mathbf{X}_j$ and $\mathbf{X}_\mathcal{G}$ in explaining $\mathbf{Y}$. One of the most important features of the CODEC is that it converges to a limit in $[0,~1]$. Given $\mathbf{X}_\mathcal{G}$, the limit is $0$ if and only if $\mathbf{Y}$ and $\mathbf{X}_j$ are conditionally independent, and is $1$ if and only if $\mathbf{Y}$ is almost surely equal to a measurable function of $\mathbf{X}_j$. If $\mathcal{G}\neq \emptyset$, let $\mathcal{G}=\{g_1,\cdots,g_q\}$, where $q\ge 1$. Let $\mathbf{x}_i^{\mathcal{G}}=(x_{ig_1},\cdots,x_{ig_q})$ be the $i$-th observation of $\mathbf{X}_\mathcal{G}$, where $i=1,\cdots,n$. CODEC is calculated as 
		\begin{eqnarray}
			T_n = T_n(\mathbf{Y},\mathbf{X}_j\mid\mathbf{X}_\mathcal{G}) = \frac {\sum_{h=1}^n {(\min\{R_h,R_{M(h)}\} - \min\{R_h,R_{N(h)} \} )}} {\sum_{h=1}^n {(R_h-\min\{R_h,R_{N(h)}\})}},\ \textrm{if}\ \mathcal{G} \ne \emptyset,
		\end{eqnarray}
		where $R_h$ denotes the rank of observation $y_h$, i.e., the number of $i$ such that $y_i \le y_h$. $M(h)$ denotes the index $i$ of the nearest neighbor $(\mathbf{x}_i^\mathcal{G}, x_{ij})$ of $(\mathbf{x}_h^\mathcal{G}, x_{hj})$ with respect to the Euclidean metric on $\mathbb{R}^{q+1}$, $N(h)$ denotes index $i$ such that $\mathbf{x}_i^\mathcal{G}$ is the nearest neighbor of $\mathbf{x}_h^\mathcal{G}$ in $\mathbb{R}^{q}$, and the ties are broken uniformly at random for both $M(h)$ and $N(h)$. $R_{M(h)}$ denotes the rank of $y_{M(h)}$, i.e., the number of $i$ such that $y_{M(i)} \le y_{M(h)}$, and $R_{N(h)}$ denotes the rank of $y_{N(h)}$.
		
		The CODEC can also measure the unconditional correlation between $\mathbf{Y}$ and a predictor $\mathbf{X}_j$, where $j\subseteq \mathcal{X}$, in the absence of any given predictors, i.e., when $\mathcal{G}=\emptyset$. In this case, the CODEC is interpreted as an unconditional dependence coefficient, and is calculated as 
		\begin{eqnarray}
			T_{n}=T_{n}(\mathbf{Y}, \mathbf{X}_j) = \frac{\sum_{h=1}^{n}\left(n \min \left\{R_{h}, R_{M(h)}\right\}-L_{h}^{2}\right)}{\sum_{h=1}^{n} L_{h}\left(n-L_{h}\right)},\ \textrm{if}\ \mathcal{G} = \emptyset,
		\end{eqnarray}
		where $R_h$ denotes the rank of $y_h$. $M(h)$ denotes the index $i$ of the nearest neighbor $x_{ij}$ of $x_{hj}$, and the ties are broken uniformly at random for $M(h)$.  $R_{M(h)}$ denotes the rank of $y_{M(h)}$. $L_h$ denotes the number of $i$ such that $y_i \ge y_h$.
	
	The calculations above require continuous $\mathbf{Y}$, $\mathbf{X}_\mathcal{G}$ and $\mathbf{X}_j$, but the CODEC can also be applied to measure the correlations between discrete predictors if the ties are unbonded randomly. If the denominator of $T_n(\mathbf{Y},\mathbf{X}_j\mid\mathbf{X}_\mathcal{G})$ is $0$, the CODEC is undefined. At this point, $Y$ is almost surely equal to a measurable function of $\boldsymbol{X}_\mathcal{G}$, and $\boldsymbol{X}_\mathcal{G}$ can be regarded as sufficient for predicting $\mathbf{Y}$.

	\subsection{The uninformative subset and its measure}
	High-dimensional data usually contains uninformative predictors because of some restrictions in data collection. Removing these predictors has limited influence on explaining the response \citep{liFeatureSelectionData2017}. For example, in the study of handwriting digits, pixels in the marginal areas may have little explanatory power and can be ignored in the construction of deep neural networks \citep{chenNonlinearVariableSelection2021}. Real datasets do not always contain uninformative predictors, since the predictors are usually carefully chosen based on expert experience before being collected to save storage space. However, if we have little prior knowledge, and obtain a dataset with a large number of uninformative predictors, removing them beforehand can greatly improve the efficiency of variable selection. 
	
	In this paper, Shannon entropy is used as the measure to distinguish the uninformative predictors from the remaining predictors. Shannon entropy is the probability of all possible values of a predictor, which represents the expectation of the amount of information contained in the predictor \citep{grayEntropy2011}. Consider a discrete predictor $\mathbf{X}_j$ that takes a finite number of $c$ possible values $x_{j,k}\in\{x_{j,1},\cdots,x_{j,c}\}$ with corresponding probabilities $p_{j,k}\in\{p_{j,1},\cdots,p_{j,c}\}$. Its entropy $H(\mathbf{X}_j)$ is defined as
	\begin{eqnarray}
		H(\mathbf{X}_j)=-\sum_{k=1}^c p_{j,k}\ln{p_{j,k}}.
	\end{eqnarray}
	In general, there is $0 \le H(\mathbf{X}_j) \le \ln{c}$. If the distribution of $\mathbf{X}_j$ is highly biased toward one of the possible value $x_{i,k}$, $H(\mathbf{X}_j)$ is the lowest. At this point, if $H(\mathbf{X}_j)=0$, $\mathbf{X}_j$ is defined as an uninformative predictor, i.e., the quantity of information contained in $\mathbf{X}_j$ is $0$. Shannon entropy can be applied only to discrete predictors, and data discretization is required beforehand for continuous predictors \citep{brownConditionalLikelihoodMaximisation2012}.
	
	\subsection{The redundant subset and its measure}
	
	If a candidate predictor and a subset of relevant predictors are collinear, keeping both of them may affect the robustness of model estimation \citep{yuEfficientFeatureSelection2004}. Although there is a consensus on the adverse effects of multicollinearity \citep{fanSelectiveOverviewVariable2010}, most variable selection methods manage to avoid discussing the issue. In this paper, such predictors are named redundant predictors, and are measured separately from other predictors to be deleted. 
	
	Gram‒Schmidt Orthogonalization (GSO) is adopted in this paper to decompose and identify the information contained in the predictor set, and further to measure multicollinearity among predictors \citep{wangUltrahighDimensionalVariable2020, lyuFilterFeatureSelection2017}. The Gram‒Schmidt theorem in Euclidean space indicates that for an index subset of predictors $\mathcal{G}=\{g_1,\cdots,g_q\}\subseteq \mathcal{X}$ and the corresponding matrix $\mathbf{X}_\mathcal{G}=(\mathbf{X}_{g_1},\cdots,\mathbf{X}_{g_q})$, if $\mathbf{X}_\mathcal{G}$ is linearly independent, one can always construct an orthogonal basis $\mathbf{Z}_\mathcal{G}=(\mathbf{Z}_{g_1},\cdots,\mathbf{Z}_{g_q})$ via GSO, where $\mathbf{Z}_\mathcal{G}$ is the linear combination of $\mathbf{X}_\mathcal{G}$, and it spans the same space as $\mathbf{X}_\mathcal{G}$.
	
	For each $\mathbf{X}_j$,\ $j\in\mathcal{G}$, the orthogonalized variable $\mathbf{Z}_j$ is
	\begin{equation} \label{eq2}
		\begin{split}
		\mathbf{Z}_{g_1} =\mathbf{X}_{g_1},\ & \\
		\mathbf{Z}_j = \mathbf{X}_j-\sum_{k=g_1}^{g_{j-1}} {\frac{\langle \mathbf{X}_j, \mathbf{Z}_k \rangle} {\Vert\mathbf{Z}_k \Vert_2^2} \mathbf{Z}_k}, & \forall j=g_2,\cdots,g_q.
		\end{split}
	\end{equation}
	where $\langle \mathbf{X}_j, \mathbf{Z}_k \rangle$ is the inner product of $\mathbf{X}_j$ and $\mathbf{Z}_k$, and $\Vert\mathbf{Z}_k \Vert_2$ is the $L^2$ norm of $\mathbf{Z}_k$.
	
	Let $q$ be the rank of $\mathbf{X}=(\mathbf{X}_1,\cdots,\mathbf{X}_p)$. Suppose $q<p$, and $\mathbf{X}_\mathcal{G}$ is linearly independent, where the index subset $\mathcal{G}\subseteq \mathcal{X}$; then, an orthogonal basis $\mathbf{Z}=\mathbf{Z}_\mathcal{G}$ of $\mathbf{X}$ can be obtained using GSO. 
	For any $j\in \mathcal{X}\ \backslash\ \mathcal{G}$, one can obtain the orthogonal variable $\mathbf{Z}_j$ of $\mathbf{X}_j$ with GSO based on $\mathbf{Z}$. If the variance of $\mathbf{Z}_j$ satisfies $Var(\mathbf{Z}_j)=0$, all the information of $\mathbf{X}_j$ is already contained in $\mathbf{X}_\mathcal{G}$, and accordingly, $\mathbf{X}_j$ is a redundant predictor, which is collinear with $\mathbf{X}_\mathcal{G}$.

	\subsection{The conditionally independent subset and its measure}
	
	In addition to uninformative and redundant predictors, there is another group of predictors that needs to be deleted. One of the most obvious features is that $\mathbf{Y}$ is conditionally independent of these predictors given the relevant subset $\mathcal{S}$. In this paper, these predictors are called conditionally independent predictors. Modeling with the conditionally independent predictors does more harm than good. Not only will it damage interpretability, but it may also reduce prediction accuracy. The CODEC can be used to measure conditionally independent predictors. In particular, given $\mathcal{S}$, if there is $T_n(\mathbf{Y}, \mathbf{X}_j\mid\mathbf{X}_\mathcal{S})$ close to $0$ for a predictor $\mathbf{X}_j$, where $j\in \mathcal{X}\ \backslash\ \mathcal{S}$, $\mathbf{X}_j$ can be considered a conditionally independent predictor.

	\section{The Proposed Transparent Variable Selection for Nonlinear and High-dimensional Data} \label{method}
	
\begin{figure}[H]
		\centering
		\includegraphics[width=0.75\linewidth]{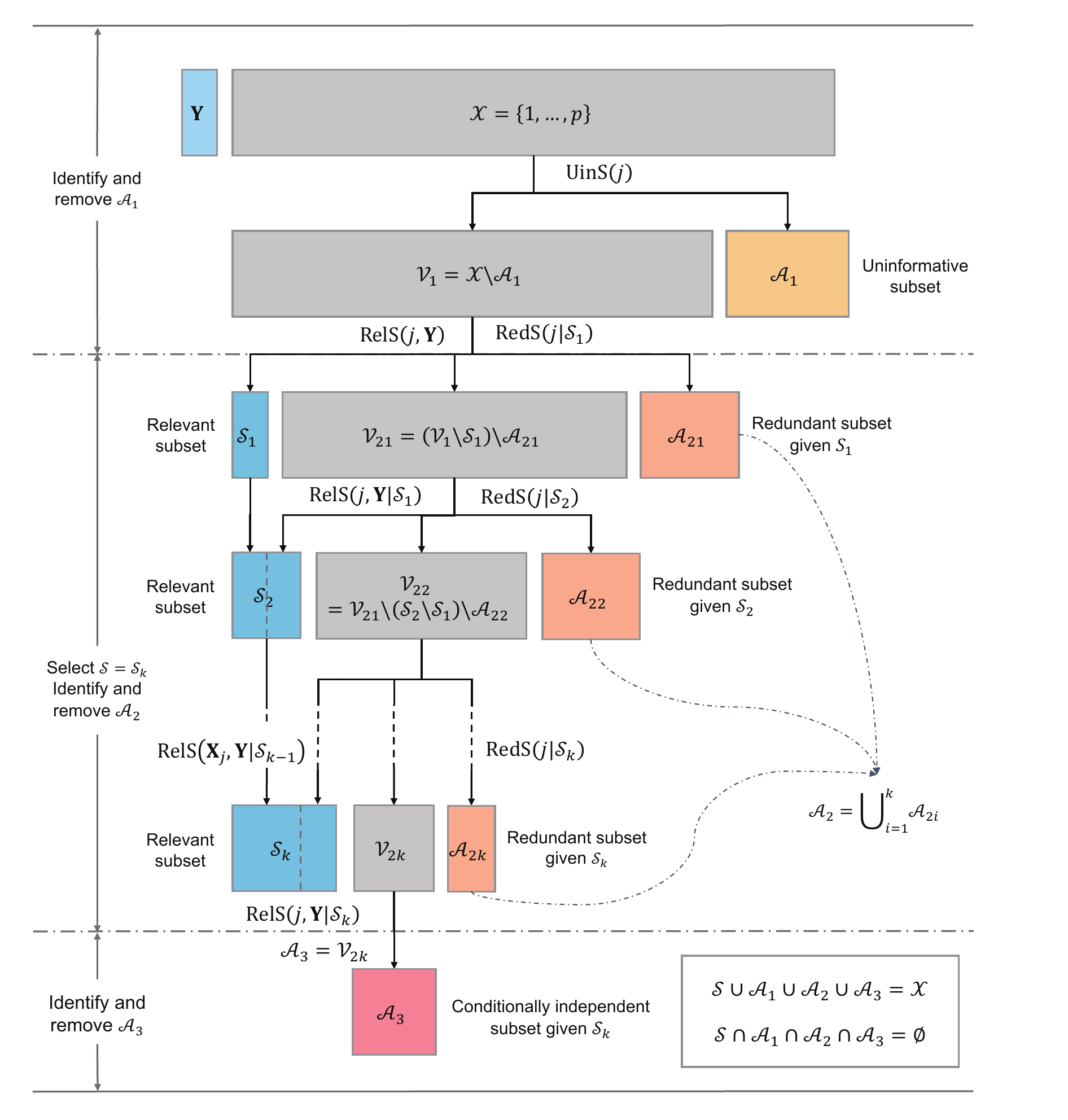}
		\caption{A flowchart of the proposed TNVS.}
		\label{fig:framework_of_TNVS}
	\end{figure}

	In this section, we propose a Transparent and Nonlinear Variable Selection (TNVS) approach for high-dimensional data. Denote $\mathcal{V}\subseteq \mathcal{X}$ as an index set of the candidate predictors, which contains the indices of all predictors that are promising to explain the response. A three-step heuristic search is established to transparently separate the candidate set into the subset to be selected and those to be deleted. The indices of the selected predictors are reserved in the relevant subset $\mathcal{S}$, and those of the removed predictors are respectively categorized into in the uninformative subset $\mathcal{A}_1$, the redundant subset $\mathcal{A}_2$, and the conditionally independent subset $\mathcal{A}_3$. A flowchart of the proposed TNVS is illustrated in Fig. \ref{fig:framework_of_TNVS}. 
	
	If we have no prior knowledge of the key variables, the initial $\mathcal{V}$ is set as the index set of input predictors $\mathcal{X}$, and the initial $\mathcal{S}$, $\mathcal{A}_1$, $\mathcal{A}_2$, and $\mathcal{A}_3$ are set as empty sets. The heuristic search of TNVS contains three steps, i.e., prefiltering, forward selection, and batch deletion. In the prefiltering step, uninformative predictors containing little information are identified and removed. Forward selection and batch deletion are iterated alternately. Every time a relevant predictor is selected, a deletion step is performed to remove redundant predictors that are collinear with all the selected predictors. 
	
	In the prefiltering step, uninformative predictors are distinguished from other predictors using an Uninformative Score (UinS), and the indices of these uninformative predictors are added to $\mathcal{A}_1$ and excluded from $\mathcal{V}$. In TNVS, the Shannon entropy of a predictor serves as the UinS.
	For an index $j\in\mathcal{V}$, if $\textrm{UinS}(j) = H(\mathbf{X}_j)<\alpha_1$ , where $\alpha_1$ is the uninformative threshold, $\mathbf{X}_j$ will be accordingly regarded as an uninformative predictor. 
	
	The forward selection step identifies the most relevant predictor to the response given $\mathcal{S}$, and places its index in $\mathcal{S}$ to form a new relevant subset. The degree of relevance between any $\mathbf{X}_j\ (j\in\mathcal{V})$ and $\mathbf{Y}$ is determined by the Relevance Score (RelS). In TNVS, the CODEC is adopted as the measure of RelS. Given $\mathcal{S}$, the RelS of any index $j$ is
	\begin{eqnarray}
		\textrm{RelS}(j,\mathbf{Y}\mid\mathcal{S}) = \left\{
		\begin{aligned} 
			& T_n(\mathbf{Y}, \mathbf{X}_j), & if\ \mathcal{S}=\emptyset \\
			& T_n(\mathbf{Y}, \mathbf{X}_j\mid\mathbf{X}_\mathcal{S}), & otherwise
		\end{aligned}
		\right.\ ,
	\end{eqnarray}
	The index with maximum $\textrm{RelS}$ is selected from $\mathcal{V}$, and the corresponding predictor is most relevant to $\mathbf{Y}$ considering its interaction with $\mathbf{X}_\mathcal{S}$. 
	
	The batch deletion step identifies multicollinearity in $\mathcal{V}$ given the relevant subset $\mathcal{S}$. Every time forward selection is performed, all redundant predictors that are collinear with the new $\mathbf{X}_\mathcal{S}$ are detected from the candidates. Their indices are added in $\mathcal{A}_2$ and removed from $\mathcal{V}$. In TNVS, GSO is adopted to identify the redundant predictors given $\mathcal{S}$. First, the orthogonal basis $\mathbf{Z}_\mathcal{S}$ of $\mathcal{S}$ is obtained. Then, GSO is performed on all candidate predictors $\mathbf{X}_j\ (j\in\mathcal{V})$ based on $\mathbf{Z}_\mathcal{S}$ to determine which candidates are collinear with $\mathbf{X}_\mathcal{S}$. The Redundancy Score (RedS) of $j$ given $\mathcal{S}$ is the variance of the orthogonalized variable $Var(\mathbf{Z}_j)$. If $\textrm{RedS}(j\mid\mathcal{S})<\alpha_3$, where $\alpha_3$ is the redundant threshold, $\mathbf{X}_j$ is regarded as a redundant predictor which is collinear with $\mathbf{X}_\mathcal{S}$.
	
		\begin{algorithm}[H]
		\small
		\caption{Pseudocode of the proposed TNVS.}
		\label{alg:algorithm framework}
		\KwData{the matrix of predictors $\mathbf{X}=(\mathbf{X}_1,\cdots,\mathbf{X}_p)$, their index set $\mathcal{X}=\{1,\cdots,p\}$, the response vector $\mathbf{Y}$}
		\KwIn{uninformative threshold $\alpha_1$, relevant threshold $\alpha_2$, redundant threshold $\alpha_3$, maximum model size $d_{\max}$ (optional)}
		\KwOut{$\mathcal{S}$ to be selected, and $\mathcal{A}_1$, $\mathcal{A}_2$, and $\mathcal{A}_3$ to be removed}
		Set the initial candidate feature subset $\mathcal{V}=\mathcal{X}$, and the initial $\mathcal{S}$, $\mathcal{A}_1$, $\mathcal{A}_2$, $\mathcal{A}_3$ as $\emptyset$\;
		\For(\tcp*[f]{Obtain the uninformative subset $\mathcal{A}_1$})
		{all $j \in \mathcal{V}$}
		{
			\lIf {$\mathrm{UinS}(j) < \alpha_1$} 
			{$\mathcal{V} =\mathcal{V} \backslash \{j\}$; $\mathcal{A}_1 =\mathcal{A}_1 \cup \{j\}$}
		}
		\While{$\mathcal{V}\ne \varnothing$ and $\lvert \mathcal{S} \rvert <d_{max}$ (if provided)}
		{
			\eIf{any $\mathrm{RelS}(j,\mathbf{Y}\mid\mathcal{S})$, $j\in \mathcal{V}$ is undefined}
			{\textbf{break}}
			{
				$k=\mathop{\arg\max}\limits_{j \in \mathcal{V}}\{\textrm{RelS}(j,\mathbf{Y}\mid\mathcal{S})\}$; \tcp*[f]{Obtain the relevant subset $\mathcal{S}$} \\
				\eIf {$\mathrm{RelS}(k,\mathbf{Y}\mid\mathcal{S})<\alpha_2$}
				{\textbf{break}}
				{$\mathcal{V}=\mathcal{V} \backslash \{k\}$; $\mathcal{S}=\mathcal{S} \cup \{k\}$\;
					Obtain $\mathbf{Z}_\mathcal{S}$\;
				}
				\For(\tcp*[f]{Obtain the redundant subset $\mathcal{A}_2$}){all $j \in \mathcal{V}$}
				{
					\lIf{$\mathrm{RedS}(j\mid\mathcal{S}) < \alpha_3$}
					{$\mathcal{V} =\mathcal{V} \backslash \{j\}$; $\mathcal{A}_2 =\mathcal{A}_2 \cup \{j\}$}
				}
			}
		}
		$\mathcal{A}_3 =\mathcal{A}_3 \cup \mathcal{V}$; \tcp*[f]{Obtain the conditionally independent subset $\mathcal{A}_3$} \
	\end{algorithm}

	The termination criterion of TNVS is the cardinality of the relevant subset $\lvert \mathcal{S} \rvert$ reaching the predetermined upper bound $d_{\max}$, or $\textrm{RelS}(j,\mathbf{Y}\mid\mathcal{S})<\alpha_2$ or undefined for all $j\in\mathcal{V}$, where $\alpha_2$ is the relevant threshold. If the termination criterion is satisfied, all remaining elements in $\mathcal{V}$ belong to the conditionally independent subset $\mathcal{A}_3$. When TNVS is finished, the original index set of predictors $\mathcal{X}$ is divided into a relevant subset $\mathcal{S}$, an uninformative subset $\mathcal{A}_1$, a redundant subset $\mathcal{A}_2$, and a conditionally independent subset $\mathcal{A}_3$, resulting in $\mathcal{S} \cup \mathcal{A}_1 \cup \mathcal{A}_2 \cup \mathcal{A}_3 = \mathcal{X}$ and $\mathcal{S} \cap \mathcal{A}_1 \cap \mathcal{A}_2 \cap \mathcal{A}_3 =\varnothing$. The pseudocode of TNVS is shown in Algorithm \ref{alg:algorithm framework}, and an example to illustrate the procedure of TNVS is provided in Appendix \ref{subsec:A2}.

	The computational time of the proposed TNVS is $O(dn\log{n})$, where $d$ is the number of selected predictors. First, the computational cost of the prefiltering step is $O(np)$. The complexity of the forward selection step is $O(np\log{n})$ since it takes $O(n\log{n})$ to calculate the CODEC of all the predictors and sort the most relevant predictor \citep{azadkiaSimpleMeasureConditional2021a}. For the batch deletion step, GSO requires $O(np)$ to screen all predictors. Forward selection and batch deletion are iterated $d$ times on average. The total cost of TNVS is $O(dpn\log{n})$. In the worst case when $d$ is proportional to $p$, the time complexity is $O(np^2\log{n})$. Suppose there are no less than $p$ processors, predictors can be measured in parallel in each step. In summary, the computational time of TNVS is $O(dn\log{n})$ on average. As is shown in simulations, the proposed TNVS is efficient compared with baselines even with limited processors.
	
	The main advantages of the presented TNVS are summarized as follows. First, a transparent framework of variable selection is achieved using a three-step heuristic search, which groups predictors into four mutually disjoint subsets. In every step of the framework, TNVS can provide reasonable interpretations to select or delete certain predictors. Second, the recently proposed CODEC is introduced in TNVS to identify the complex nonlinear associations between the response and the predictors. TNVS can handle not only high-dimensional data in which the response is monotonic and additive functionally dependent on the predictors, but also those data with oscillatory functional dependence and interactions. Moreover, TNVS is model-free. The selected subset of predictors has strong generality and can be used to construct many kinds of learning models.

	\section{Simulation studies} \label{Simulations}
	Substantial experiments on simulated datasets are implemented to illustrate the effectiveness and interpretability of TNVS. First, the variable selection results of TNVS are compared with several competitive baselines to show its validity and conciseness. Then, the subsets of predictors are further evaluated to demonstrate model interpretability.
	
	\subsection{The simulated datasets}
	We generate a regression simulation problem with both monotonic and nonmonotonic functional dependence, interactions, uninformative predictors, multicollinearity, and conditionally independent predictors. In the problem, $90\%$ of the $p$-dimensional input predictors have information, and the remaining $10\%$ are uninformative. The first $90\%$ predictors are consisted of 9 signals $\mathbf{X}_{t_1},\cdots,\mathbf{X}_{t_9}$, where $t_g=(g-1)\cdot p/10+1,\ g=1,\cdots,9$, and the rest are redundant predictors that are collinear with these $9$ signals. The signals are independently sampled from the standard normal distribution $\mathcal{N}(0,~1)$. The redundant predictors are highly correlated with one of the signals, and there is $\mathbf{X}_{t_g+j}=\mathbf{X}_{t_g}+\lambda\mathbf{\epsilon}_{t_g+j},\ g=1,\cdots,9,\ j=1,\cdots,(p/10)$, where $\mathbf{\epsilon}_{t_g+j}$ is the stochastic error that follows $\mathcal{N}(0,~1)$, and $\lambda$ is set as $0.01$. The last $10\%$ predictors are uninformative with only a few nonzero observations. The proportion of nonzero observations is set as $0.001$, and nonzero observations are randomly generated from $\mathcal{N}(0,~ 0.1^2)$. The response $\mathbf{Y}$ is a nonlinear measurable function of some predictors
	\begin{eqnarray}
		\mathbf{Y} = 2\mathbf{X}_{t_1}\mathbf{X}_{t_2} + \cos(\pi \mathbf{X}_{t_3}\mathbf{X}_{t_4}) + \mathbf{\varepsilon},
	\end{eqnarray}
	where $\mathbf{\varepsilon}$ is the stochastic error that follows $\mathcal{N}(0,~0.1^2)$. 
	$\mathbf{Y}$ is nonlinearly correlated with these four predictors or any of their collinear predictors. $\mathbf{Y}$ is oscillatory functionally dependent on $\mathbf{X}_{t_3}$ and $\mathbf{X}_{t_4}$, and there are interactions between $\mathbf{X}_{t_1}$ and $\mathbf{X}_{t_2}$, and $\mathbf{X}_{t_3}$ and $\mathbf{X}_{t_4}$. 
	The problem has the following three settings by specifying $n$ and $p$. For each setting, $10$ datasets are generated. 
	\begin{itemize}
		\item[] Setting 1:\ $n=2\ 000,~p=1\ 000$;
		\item[] Setting 2:\ $n=2\ 000,~p=2\ 000$;
		\item[] Setting 3:\ $n=2\ 000,~p=5\ 000$.
	\end{itemize}

	\subsection{Baselines and parameter settings} \label{subsection:benchmarking}
	In our experiments, TNVS is evaluated against six representative baselines, including two up-to-date model-free and nonlinear methods, namely FOCI \citep{azadkiaSimpleMeasureConditional2021a} and WLS \citep{zhongModelfreeVariableScreening2021}, two classic model-free and nonlinear feature screening methods, namely DC-SIS \citep{liFeatureScreeningDistance2012} and SIRS \citep{zhuModelFreeFeatureScreening2011}, and two stepwise methods based on linear partial correlation, namely PC-simple \citep{buhlmannVariableSelectionHighdimensional2010} and TPC \citep{liVariableSelectionPartial2017}.
	
	The parameter settings of the variable selection methods are listed as follows. For TNVS, the uninformative threshold $\alpha_1$, relevant threshold $\alpha_2$, and redundant threshold $\alpha_3$ are empirically set as $0.01,\ -0.01,\ 0.01$, as the default setting in line with previous research \citep{liuFunctionalVariableSelection2018}. $d_{\max}$ is set as $\lceil n/\log{n} \rceil$, which represents the smallest positive integer no less than $n/\log{n}$. The termination criterion of FOCI is that the CODECs of all candidates are no larger than 0, or the number of selected predictors reaches $\lceil n/\log{n} \rceil$. The significance levels of PC-simple and TPC are set as $0.05$ \citep{buhlmannVariableSelectionHighdimensional2010, liVariableSelectionPartial2017}. The number of selected predictors of WLS, DC-SIS and SIRS are empirically set as $\lceil n/\log{n} \rceil$ in line with the literature. Considering the computing capability, we set the upper bound of the CPU time as $3\ 600$ seconds for all stepwise methods, including TNVS, FOCI, PC-Simple and TPC. In addition, predictors with zero variance are removed beforehand for FOCI. If a selected subset is empty, i.e., no predictor is considered to be correlated with the response, the mean of the response is used as the predictive value for regression problems, and a random category of the response is chosen as the predictive value for classification problems. 
	
	10-fold cross-validation is performed on each dataset, where the $n$ samples of the dataset are randomly divided into $10$ equal folds. Each unique fold of samples is used for testing and the remaining $9n/10$ samples are used for training. The performance in each setting is the average result of all folds on all the $10$ corresponding datasets, i.e., each variable selection method is tested for $100$ times in each setting to evaluate their variable selection capability. 
	
	All the variable selection methods are implemented in R 4.0.3 for a fair comparison\footnote{Our code is available at https://github.com/kywang95/TNVS.}. The experiments are performed using an Intel Core i7 3.4 GHz processor with 16 GB of RAM.
	
	\subsection{Effectiveness and efficiency of the proposed method on nonlinear simulations}
	The following four criteria are adopted to evaluate the effectiveness and efficiency of TNVS and baselines on high-dimensional simulations with complex nonlinear functional dependencies. The results of each simulation in $100$ repetitions are summarized in Table \ref{tab:simu_evaluation_1}. 
	\begin{enumerate}[(1)]
		\item $P_a$: probability of the discovery of all truly important predictors over all repetitions. The closer $P_a$ is to $1$, the more robust the variable selection method is to retain all true predictors. For simulations, true predictors are any of the relevant predictors and the redundant predictors that are collinear with the relvant predictors. For example, for Setting $1$, the indices of true predictors are any combinations of four elements from the subsets $\{1,~2,\cdots,100\}$, $\{101,~102,\cdots,200\}$, $\{201,~202,\cdots, 300\}$, and $\{301,~302,\cdots, 400\}$, and $P_a$ is the frequency that contains at least one element from each of the four sets.
		\item $\mathcal{M}$: the minimum model size required to include all true predictors in all repetitions in which none of the true predictors are omitted. The closer $\mathcal{M}$ is to the true model size, the more concise the selected subset is. The average and the standard deviation of $\mathcal{M}$ are obtained.
		\item Coverage: The number of true predictors covered by the selected subset. Coverage is less than or equal to the number of true predictors. The average and standard deviation of the coverage in all repetitions are obtained. In the simulated datasets, the closer the coverage is to $4$, the fewer true predictors are omitted. 
		\item Time: the average and standard deviation of running time (in seconds) of variable selection in $100$ repetitions.
	\end{enumerate}
		
	\begin{table}[htbp]
		\centering
		\small
		\caption{$P_a$, the average and the standard deviation (in parentheses) of $\mathcal{M}$, coverage, and running time (in seconds) in $100$ repetitions. The best results are presented in bold, and the second-best results are underlined.}
		\begin{tabular}{lcccccccc}
			\toprule
			Setting & Index & TNVS & FOCI & WLS & SIRS & DC-SIS & PC-simple & TPC \\
			\midrule
			\multirow{4}{*}{1} & $P_a$ & $\underline{0.99}$ & 0.98 & $\mathbf{1.00}$ & 0.00 & 0.02 & 0.00 & 0.00 \\
			\specialrule{0em}{1pt}{1pt}
			& $\mathcal{M}$ &  $\underset{(0.00)}{\mathbf{4.00}}$ & $\underset{(0.10)}{\mathbf{\underline{4.01}}}$ & $\underset{(9.75)}{26.28}$   & $\underset{(-)}{-}$ & $\underset{(10.61)}{223.50}$ & $\underset{(-)}{-}$ & $\underset{(-)}{-}$ \\
			& Coverage & $\underset{(0.20)}{\underline{3.98}}$ & $\underset{(0.28)}{3.96}$  & $\underset{(0.00)}{\mathbf{4.00}}$ & $\underset{(0.60)}{1.11}$ & $\underset{(0.14)}{3.02}$  & $\underset{(0.54)}{0.50}$  & $\underset{(0.00)}{0.00}$  \\
			& Time(s) & $\underset{(1.98)}{33.45}$ & $\underset{(127.05)}{598.84}$ & $\underset{(0.72)}{90.43}$ & $\underset{(4.25)}{10.27}$ & $\underset{(3.42)}{245.74}$ & $\underset{(0.62)}{4.41}$ & $\underset{(0.16)}{4.41}$ \\
			\midrule
			\multirow{4}{*}{2} & $P_a$ & $\underline{0.97}$ & 0.93 & $\mathbf{1.00}$ & 0.00 & 0.00 & 0.00 & 0.00 \\
			\specialrule{0em}{1pt}{1pt}
			& $\mathcal{M}$ & $\underset{(0.00)}{\mathbf{4.00}}$ & $\underset{(0.10)}{\underline{4.01}}$ & $\underset{(9.10)}{28.59}$ & $\underset{(-)}{-}$ & $\underset{(-)}{-}$ & $\underset{(-)}{-}$ & $\underset{(-)}{-}$ \\
			& Coverage & $\underset{(0.49)}{\underline{3.92}}$ & $\underset{(0.61)}{3.84}$  & $\underset{(0.00)}{\mathbf{4.00}}$ & $\underset{(0.59)}{0.72}$ & $\underset{(0.00)}{2.00}$ & $\underset{(0.53)}{0.31}$ & $\underset{(0.00)}{0.00}$ \\
			& Time(s) & $\underset{(6.64)}{68.69}$ & $\underset{(503.19)}{2826.55}$ & $\underset{(3.89)}{530.74}$ & $\underset{(7.62)}{28.41}$ & $\underset{(19.85)}{527.68}$ & $\underset{(55.14)}{30.69}$ & $\underset{(0.27)}{11.11}$ \\
			\midrule
			\multirow{4}{*}{3} & $P_a$ & $\underline{0.99}$ & 0.97 & $\mathbf{1.00}$ & 0.00 & 0.00 & 0.00 & 0.00 \\
			\specialrule{0em}{1pt}{1pt}
			& $\mathcal{M}$ & $\underset{(0.00)}{\mathbf{4.00}}$ & $\underset{(0.00)}{\mathbf{4.00}}$ & $\underset{(8.87)}{33.49}$ & $\underset{(-)}{-}$ & $\underset{(-)}{-}$ & $\underset{(-)}{-}$ & $\underset{(-)}{-}$ \\
			& Coverage & $\underset{(0.20)}{\underline{3.98}}$ & $\underset{(0.34)}{3.94}$ & $\underset{(0.00)}{\mathbf{4.00}}$ & $\underset{(0.50)}{0.74}$ & $\underset{(0.14)}{1.02}$ & $\underset{(0.55)}{0.28}$ & $\underset{(0.00)}{0.00}$ \\
			& Time(s) & $\underset{(9.97)}{236.80}$ & $\underset{(36.46)}{3670.26}$ & $\underset{(3.73)}{1300.66}$ & $\underset{(1.57)}{96.59}$ & $\underset{(59.23)}{1321.23}$ & $\underset{(2308.58)}{4908.73}$ & $\underset{(0.89)}{50.60}$ \\
			\bottomrule
		\end{tabular}%
		\label{tab:simu_evaluation_1}
	\end{table}
	
	Table \ref{tab:simu_evaluation_1} shows that our TNVS, and the recently proposed FOCI and WLS, outperform the other traditional methods, and TNVS is the most effective and efficient one among these three methods. First, the-close to-1 $P_a$ in all settings indicates that these methods can reserve all true predictors in most cases. Among these three methods, TNVS has the smallest minimum model size, which means that the subset selected by TNVS is the most concise. In addition, computational time of TNVS is the least among these three methods, which shows that it is the most efficient. The framework of TNVS is most similar to that of the FOCI. Comparing these two methods, TNVS has a smaller minimum model size and a larger coverage, indicating that TNVS is more accurate and reliable than FOCI.
	
	Table \ref{tab:simu_evaluation_1} also shows that these nonlinear simulation problems are extremely difficult to solve with the other four prevailing methods, i.e., SIRS, DC-SIS, PC-simple and TPC. Their $P_a$ are close to 0, indicating that these methods omit some true predictors and are invalid. PC-simple and TPC are designed for linear regression models, and close-to-0 coverages indicate their weak abilities to identify nonlinearly relevant predictors. SIRS and DC-SIS can only identify the monotonic nonlinear relevance, and having difficulty in removing the multicollinearity, and close-to-0 coverages indicate they cannot retain all important predictors.
	
	Experiments on simulations demonstrate that every module of the proposed TNVS is effective and indispensable. First, compared with PC-simple and TPC, these two stepwise methods with linear partial correlation, TNVS adopts the nonlinear CODEC. This leads to  the unsatisfying $P_a$ of these two baselines and the close-to-1 $P_a$ of TNVS, which indicates that introducing CODEC ensures that TNVS can effectively identify the complex nonlinear associations, such as interacted predictors, and oscillatory functional dependence with interactions. Second, FOCI is a stepwise method with CODEC, but ignoring to sorted disposal the uninformative and redundant predictors. Compared with FOCI, TNVS avoids retaining the large percentages of these unnecessary predictors, and achieves larger $P_a$, smaller model size, larger coverage, and less computational time. This implies that prefiltering and batch deletion inherently improve effectiveness and efficiency of TNVS.

	\subsection{Interpretability of the proposed method}
	We further demonstrate the model interpretability of the proposed variable selection method. Since in simulations, the truly relevant, uninformative, redundant, and conditionally independent predictors are known in advance, we can obtain the proportions of four types of predictors in the selected subset for all methods.
	Taking a closer look at the three competitive methods mentioned in Table \ref{tab:simu_evaluation_1}, Table \ref{tab:simu_evaluation_2} shows the proportions of the ground truth relevant ($\textrm{Rel}_{GT}$), uninformative ($\textrm{Uin}_{GT}$), redundant ($\textrm{Red}_{GT}$), and conditionally independent ($\textrm{Cind}_{GT}$) predictors in the selected subset of TNVS, FOCI, and WLS. The average of each quantity in all $100$ repetitions are summarized. The truly relevant predictors in the selected subset are true positive, whereas the rest three types are false positive. Thus, $\textrm{Rel}_{GT}$ is actually the precision of each methods. The larger $\textrm{Rel}_{GT}$ is, and the smaller the proportions of $\textrm{Rel}_{GT}$, $\textrm{Uin}_{GT}$, and $\textrm{Red}_{GT}$ are, the more concise the selected subset is.

		\begin{table}[htbp]
		\centering
		\small 
		\caption{Average proportions of four types of predictors in the selected subsets of predictors obtained by TNVS, FOCI, and WLS. The best results are presented in bold, and the second-best results are underlined.}
		\begin{tabu}{lcccc}
			\tabucline[0.08em]{-}
			Setting & Index & TNVS  & FOCI & WLS \\
			\hline
			\multicolumn{1}{l}{\multirow{4}[1]{*}{1}} & $\textrm{Rel}_{GT}$ & $\mathbf{1.00}$ & $\underline{0.12}$ & 0.02 \\
			&  $\textrm{Uin}_{GT}$ & $\mathbf{0.00}$ & 0.83 & $\underline{0.03}$ \\
			& $\textrm{Red}_{GT}$  & $\mathbf{0.00}$ & $\underline{0.05}$ & 0.42 \\
			& $\textrm{Cind}_{GT}$ & $\mathbf{0.00}$ & $\mathbf{0.00}$  & 0.53 \\
			\hline
			\multicolumn{1}{l}{\multirow{4}[1]{*}{2}} & $\textrm{Rel}_{GT}$ & $\mathbf{0.99}$ & $\underline{0.06}$ & 0.02 \\
			& $\textrm{Uin}_{GT}$ & $\mathbf{0.00}$ & 0.92 & $\underline{0.05}$ \\
			& $\textrm{Red}_{GT}$ & $\mathbf{0.00}$ & $\underline{0.02}$ & 0.42 \\
			& $\textrm{Cind}_{GT}$ & $\mathbf{0.01}$ & $\underline{0.00}$ & 0.51 \\
			\hline
			\multicolumn{1}{l}{\multirow{4}[1]{*}{3}} & $\textrm{Rel}_{GT}$ & $\mathbf{1.00}$ & $\underline{0.10}$ & 0.02 \\
			& $\textrm{Uin}_{GT}$ & $\mathbf{0.00}$  & 0.87 & $\underline{0.07}$ \\
			& $\textrm{Red}_{GT}$ & $\mathbf{0.00}$  & $\underline{0.03}$  & 0.41 \\
			& $\textrm{Cind}_{GT}$ & $\mathbf{0.00}$ & $\mathbf{0.00}$ & 0.50 \\
			\tabucline[0.08em]{-}
		\end{tabu}%
		\label{tab:simu_evaluation_2}
	\end{table}
	Table \ref{tab:simu_evaluation_2} shows that TNVS has the largest proportion of $\textrm{Red}_{GT}$ in its selected subsets, or equivalently, the largest precision. Nearly all the predictors selected by TNVS are relevant predictors. Uninformative, redundant, and conditionally independent predictors are rarely included in the selected subset of TNVS. The subset selected by FOCI contains a large percentage of uninformative predictors and some redundant predictors. Although WLS can select all relevant predictors, they only take a small part of its selected subset, and many unimportant predictors are falsely picked before the truly important predictors. These findings demonstrate that TNVS enhances the capability of FOCI to handle the lack of information and multicollinearity, and it also achieves a better balance between accuracy and conciseness than WLS.
		
	The most significant difference of TNVS against other methods is that TNVS can divide the predictors into four disjoint subsets, instead of only the selected and the deleted ones. Table \ref{tab:simu_evaluation_3} shows the proportions of the ground truth types of predictors in the four subsets obtained by TNVS (the relevant subset $\textrm{Rel}_{pred}$, the uninformative subset $\textrm{Uin}_{pred}$, the redundant subset $\textrm{Red}_{pred}$, and the conditionally independent subset $\textrm{Cind}_{pred}$). The average of each quantity in all $100$ repetitions are summarized. 
	The truly relevant predictors grouped into the selected relevant subset are true positive, whereas those grouped into the other three types are false negative. Thus, elements in the diagonal are actually the recall of TNVS. The larger elements in the diagonal are, and the smaller the other elements are, the more accurate the variable selection is.

			\begin{table}[htbp]
		\centering
		\small 
		\caption{Average proportions of each ground truth type of predictors in the four subsets divided by TNVS ($\textrm{Rel}_{pred}$, $\textrm{Uin}_{pred}$, $\textrm{Red}_{pred}$, and $\textrm{Cind}_{pred}$).}
		\begin{tabu}{lccccc}
			\tabucline[0.08em]{-}
			Setting & Index & $\textrm{Rel}_{Pred}$ & $\textrm{Uin}_{pred}$ & $\textrm{Red}_{pred}$  & $\textrm{Cind}_{pred}$ \\
			\hline
			\multicolumn{1}{l}{\multirow{4}[1]{*}{1}} & $\textrm{Rel}_{GT}$ & 0.995 & 0 & 0 & 0.005  \\
			&  $\textrm{Uin}_{GT}$ & 0 & 1 & 0 & 0 \\
			& $\textrm{Red}_{GT}$  & 0 & 0 & 0.995 & 0.005 \\
			& $\textrm{Cind}_{GT}$ & 0 & 0  & 0  & 1 \\
			\hline
			\multicolumn{1}{l}{\multirow{4}[1]{*}{2}} & $\textrm{Rel}_{GT}$ & 0.980 & 0 & 0 & 0.020 \\
			& $\textrm{Uin}_{GT}$ & 0 & 1 & 0 & 0 \\
			& $\textrm{Red}_{GT}$ & 0 & 0 & 0.980 & 0.020 \\
			& $\textrm{Cind}_{GT}$ & 2e-5 & 0 & 3.98e-3 & 0.996 \\
			\hline
			\multicolumn{1}{l}{\multirow{4}[1]{*}{3}} & $\textrm{Rel}_{GT}$ & 0.995 & 0 & 0 & 0.005 \\
			& $\textrm{Uin}_{GT}$ & 0  & 1 & 0 & 0 \\
			& $\textrm{Red}_{GT}$ & 0  & 0  & 0.995 & 0.005  \\
			& $\textrm{Cind}_{GT}$ & 0 & 0 & 0 & 1 \\
			\tabucline[0.08em]{-}
		\end{tabu}%
		\label{tab:simu_evaluation_3}
	\end{table}
	
	Table \ref{tab:simu_evaluation_3} demonstrates the unique strength of TNVS. Besides the selected relevant subset, it further divides the predictors to be removed into the uninformative, redundant, and conditionally independent subsets. The diagonal elements show that the recall of TNVS is high on all four subsets. All uninformative predictors are correctly deleted. A very few relevant predictors are falsely identified as conditionally independent in rare cases, and the misclassification of relevant predictors hinders TNVS from distinguishing the redundancy and conditionally independence. Sufficient evidence shows that when tackling simulations with complex nonlinear functional dependencies, uninformative predictors and collinearity, TNVS provides credible explanations on why certain predictors are removed. 
	To sum up, TNVS outperforms the baselines on model interpretability.

	\section{Real data applications} \label{Realdata}
	
		\subsection{Datasets}
	Four prevailing real datasets from various domains are chosen to demonstrate the performance of the proposed TNVS \citep{liFeatureSelectionData2017}. The descriptive statistics of the datasets are listed in Table \ref{tab:table_parameter_of_datasets}. In this paper, the original probes, features, or pixels are directly used as input predictors without any preprocessing such as feature extraction, since the aim of these experiments is to demonstrate the predictive capability and interpretability of the proposed variable selection method in high-dimensional data, not to design a distinct method for specific domain, such as face images recognition.
	
	\begin{table}[htbp]
		\centering
		\small
		\caption{The summary statistics of the real datasets employed to test the competence of the variable selection methods, including abbreviated names of these datasets, numbers of categories $c$, sample sizes $n$, numbers of input predictors $p$, and brief descriptions.}
		\begin{tabular}{lcccl}
			\toprule
			Name & $c$ & $n$ & $p$ & Description \\
			\midrule
			arcene & 2 & 200 & 10\ 000 & mass spectrometry \\
			isolet & 6 & 300 & 617 & voices of spoken letters \\
			warpAR10P & 10 & 240 & 2\ 400 & face images \\
			ORL & 40 & 400 & 4\ 096 & face images \\
			\bottomrule
		\end{tabular}
		\label{tab:table_parameter_of_datasets}
	\end{table}
	
	The reasons for choosing these datasets are as follows. First, these biological or image datasets are high-dimensional data with a large number of predictors. The response and predictors are often nonlinearly correlated in such datasets, and the inputs usually contain a large number of uninformative, redundant, or conditionally independent predictors. Moreover, the face images are considered as the preference since we can easily visualize the variable selection results on the original images, and literature on facial landmark detection \citep{kostingerAnnotatedFacialLandmarks2011} has provided some key reference points. By checking the relevance between our selected predictors and these key reference points, we can easily observe whether the selection results have evident post hoc interpretability. Two ORL datasets with different resolutions are selected to examine whether there are more redundant predictors in images with higher resolution, i.e., whether the images with lower resolution is distinct enough for classification. The datasets are described in detail.
	\begin{enumerate}[(1)]
		\item arcene is a binary classification dataset to discriminate between patients with and without cancer. The data are collected from mass spectrometry analysis of blood serum, and the predictors are continuous. The sample contains $88$ observations with ovarian or prostate cancer and $112$ observations of healthy patients. Each observation has $7\ 000$ real features and $3\ 000$ random probes. These random probes are permuted least-informative feats, and some of them can probably be identified as uninformative predictors.
		\item isolet is an audio dataset for predicting what letter name is spoken. The dataset contains $1\ 560$ observations, which are audio recordings of $30$ speakers. Each subject speaks names of $26$ letters of the alphabet twice. Each observation contains $617$ features. In this experiment, we generate a high-dimensional dataset with $300$ observations from the raw isolet: the first $6$ categories are considered, and $50$ observations are randomly sampled from each category.
		\item warpAR10P is a facial recognition dataset with 10 distinct subjects. For each subject, $13$ images are captured under various lighting conditions, with different facial expressions and occlusions such as scarfs and sunglasses. The size of each image is $60\times 40$ pixels, with $256$ gray levels per pixel.
		\item ORL is another facial recognition dataset containing $400$ face images collected from $40$ individuals, and each image contains $64\times 64$ pixels with $256$ gray levels per pixel. These images exhibit multiple changes, such as various picture-taking times, illumination changes, different expressions, and facial detail changes.
	\end{enumerate}
	
	Note that Z-score normalization is adopted for input predictors in all datasets, i.e., $\tilde{\mathbf{X}}_j = (\mathbf{X}_j – \mu_j) / \sigma_j$, where $\mu_j$ and $\sigma_j$ are the mean and standard deviation of predictor $\mathbf{X}_j$, respectively. 
	
\subsection{Learning models and parameter settings} 
	The proposed TNVS is inherently model-free and the selection process relies on no assumptions of the learning model. The effectiveness and generality of the selected subsets are consequently evaluated in terms of their performance on multiple learning models. Specifically, five prevailing predictive models commonly adopted in literature are considered \citep{salesiTAGATabuAsexual2021, wanR2CIInformationTheoreticguided2022}, including Support Vector Regression (SVR) / Support Vector Machine (SVM), Random Forest (RF), Decision Tree (DT), Light Gradient Boosting Machine (lightGBM) and Multi-layer Perceptron (MLP). To examine the predictive capabilities of these variable selection methods, each learning model is trained with the selected predictors, and the results are then compared. These experiments aim to demonstrate the superior generality of variable selection methods rather than the enhanced prediction performance of well-tuned models. Thus, the configurations of the learning models are set to be identical for all variable selection methods to avoid favoring any individual method by combining it with more competitive models. Most parameters are kept as the default settings in scikit-learn \citep{sklearn_api}. The only adjustment is that the number of iterations is set as $5\ 000$ for MLP. All the learning models are implemented in Python 3.9 using the application programming interface (API) of scikit-learn.
	
	For real datasets, repeated $10$-fold cross-validation is adopted to evaluate each combination of variable selection methods and learning models. The $10$-fold cross-validation is repeated $10$ times on each dataset. All variable selection methods and learning models share the same cross-validation splits so that unbiased estimates are obtained. For all splits, the ratio of classes in the training set is kept the same as that in the full sample. The predictive results are evaluated with out-of-sample accuracy (ACC), recall, $F_1$, and the Cohen Kappa Score (Kappa). The ranges of ACC, recall and $F_1$ are within $[0,~1]$; the larger these indicators are, the better the prediction is. The range of Kappa is within $[-1,~1]$; the closer the score is to 1, the better the prediction is.
	
	\subsection{Effectiveness on real datasets}
	
\begin{figure}[H]
		\centering
		\includegraphics[width=0.85\linewidth]{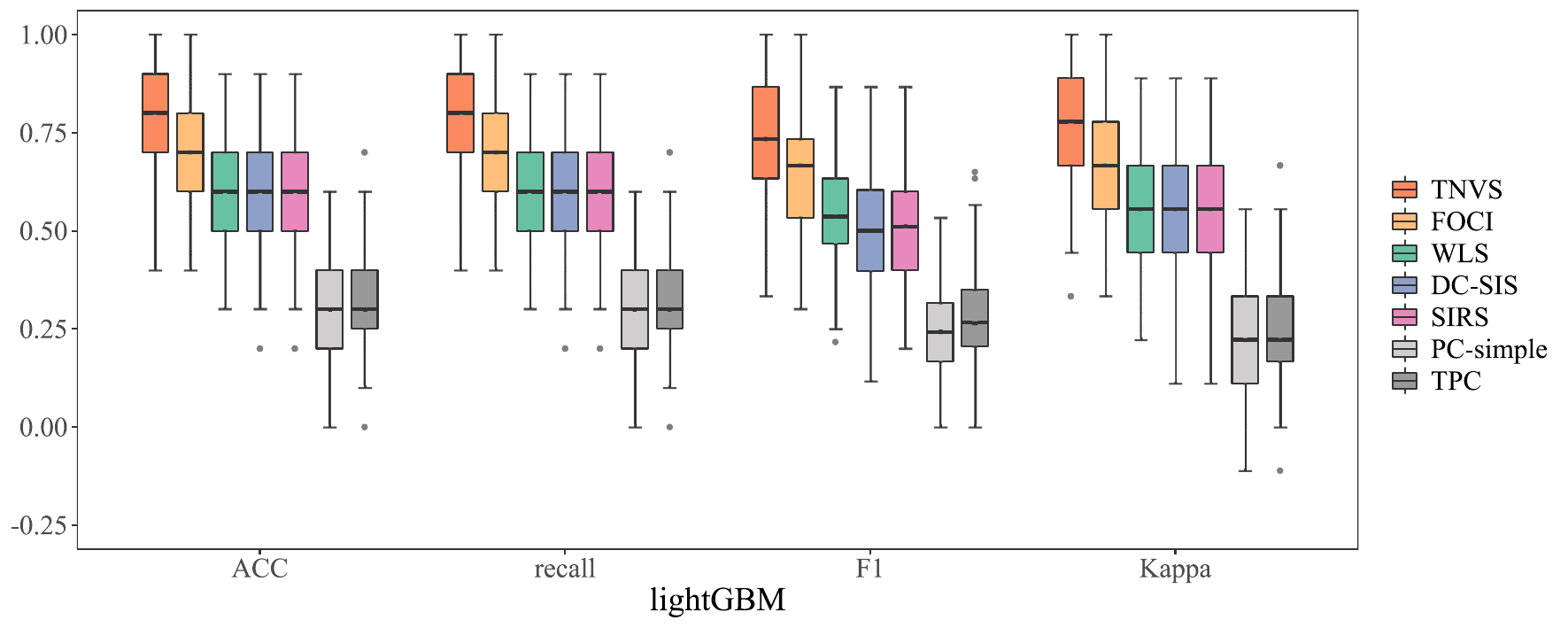}
		\caption{A boxplot of the average value of indicators in $100$ repetitions on warpAR10P, which is obtained by lightGBM trained with the selected predictors obtained using different variable selection methods.}
		\label{fig:real_prediction}
	\end{figure}

	Since the true models of real datasets are unknown, the predictive results of selected subsets are used to indirectly demonstrate the effectiveness of variable selection.
	The predictive results obtained using multiple combinations of variable selections and learning models are demonstrated as follows. First, Fig. \ref{fig:real_prediction} shows the boxplot of the four predictive indicators obtained with lightGBM on warpAR10P, which compares the predictive capability of the proposed TNVS with the baselines in this case. The predictive results of TNVS are better than those of the baselines, especially PC-Simple and TPC. Other statistical results of the average ACC, recall, $F_1$, and Kappa on the real datasets are listed in Table \ref{tab:appendix_A1} of \nameref{appendix_1}. The results of other datasets and other learning models substantially agree with those in Fig. \ref{fig:real_prediction}, which demonstrates that TNVS is robustly superior to the baselines. 
	
	A statistical test is performed to demonstrate the predictive capability of the proposed TNVS. For every dataset, the Friedman test is performed on statistical results of the four indicators obtained by all learning models to determine the overall performance of a certain variable selection method. The Friedman test provides a mean rank of these methods, and indicates whether there are significant differences among these methods. A higher mean rank means that the method has a better-ranked performance over the competitors. Table \ref{tab:prediction_on_real_datasets} shows the result of Friedman test, including the mean rank of these methods, the order of the mean rank, and the p-value of the Friedman test. As seen, the predictors selected by TNVS attain the top two among the seven methods on real datasets under the condition of retaining less predictors than feature screening methods, and there are significant differences among these methods, which further confirms the strength and robustness of TNVS in prediction enhancement. 
	
\begin{table}[htbp]
	\centering
	\small
	\caption{The Friedman test of statistical results obtained on real datasets. The best results are presented in bold, and the second-best results are underlined.}
	
	\begin{tabular}{lccccccccc}
		
		\toprule
		Dataset & Indicator & TNVS & FOCI & WLS & SIRS & DC-SIS & PC-simple & TPC & $p$-value \\
		\midrule
		\multicolumn{1}{l}{\multirow{2}[0]{*}{arcene}} & Mean rank & $\underline{5.80}$ & $\mathbf{6.20}$ & 2.60 & 4.20 & 4.55 & 2.90 & 1.75 & \multicolumn{1}{r}{0.000} \\
		& Order & $\underline{2}$ & $\mathbf{1}$ & 6 & 4 & 3 & 5 & 7 & \\
		\multicolumn{1}{l}{\multirow{2}[0]{*}{isolet}} & Mean rank & $\mathbf{6.55}$ & 5 & $\underline{6.45}$ & 3.00 & 4.00 & 2.00 & 1.00 & \multicolumn{1}{r}{0.000} \\
		& Order & $\mathbf{1}$ & 3 & $\underline{2}$ & 5 & 4 & 6 & 7 & \\
		\multicolumn{1}{l}{\multirow{2}[0]{*}{warpAR10P}} & Mean rank & $\mathbf{7.00}$ & $\underline{6.00}$ & 4.20 & 3.65 & 4.15 & 1.00 & 2.00 & \multicolumn{1}{r}{0.000} \\
		& Order & $\mathbf{1}$ & $\underline{2}$ & 3 & 5 & 4 & 7 & 6 & \\
		\multicolumn{1}{l}{\multirow{2}[0]{*}{ORL}} & Mean rank & $\underline{6.20}$ & 5.20 & $\mathbf{6.60}$ & 3.40 & 3.60 & 1.65 & 1.35 & \multicolumn{1}{r}{0.000} \\
		& Order & $\underline{2}$ & 3 & $\mathbf{1}$ & 5 & 4 & 6 & 7 & \\
		\bottomrule
	\end{tabular}
	
	\label{tab:prediction_on_real_datasets}
	
\end{table}

	\subsection{Interpretability on real datasets}
	
	TNVS transparently groups the predictors into four subsets, i.e., relevant, uninformative, redundant, and conditionally independent, which improves the model interpretability of TNVS. The average cardinalities of the four subsets are counted to check whether these types of predictors exist in the real datasets. Besides, the average compression rate is calculated, which is defined as the proportion of relevant predictors in the inputs. The results of $100$ repetitions on each dataset are shown in Table \ref{tab:real_feature_selection}.
	
		\begin{table}[htbp]
		\centering
		\small
		\caption{The average numbers of relevant, uninformative, redundant, and conditionally independent predictors obtained by TNVS on real datasets and their standard deviation (in parentheses), and the mean compression rates achieved by TNVS.}
		\begin{tabular}{lccccc}
			\toprule
			Dataset & Relevant & \makecell[c]{Compression rate} & Uninformative & Redundant & \makecell[c]{Conditionally independent} \\ 
			\midrule
			arcene & $\underset{(10.53)}{24.35}$ & 0.24\% & $\underset{(5.95)}{45.72}$ & $\underset{(62.00)}{95.91}$ & $\underset{(71.71)}{9834.02}$ \\
			isolet & $\underset{(11.55)}{35.81}$ & 5.80\% & $\underset{(0.00)}{2.00}$ & $\underset{(3.69)}{7.21}$ & $\underset{(12.66)}{571.98}$ \\
			warpAR10P & $\underset{(7.25)}{18.29}$ & 0.76\% & $\underset{(0.00)}{0.00}$ & $\underset{(5.61)}{5.10}$ & $\underset{(11.25)}{2376.61}$ \\
			ORL & $\underset{(21.72)}{37.18}$ & 0.91\% & $\underset{(0.00)}{0.00}$ & $\underset{(21.49)}{27.95}$ & $\underset{(38.24)}{4030.87}$ \\
			\bottomrule
		\end{tabular}
		\label{tab:real_feature_selection}
	\end{table}
	
	As shown in Table \ref{tab:real_feature_selection}, the percentage of relevant predictors is lower than $6\%$ in all datasets, and is even lower than $1\%$ in arcene, warpAR10P, and ORL, which indicates that TNVS compresses the input set of predictors to a large degree. Redundant predictors are identified in all datasets, which shows the usefulness of GSO. TNVS identifies the uninformative predictors in arcene as expected.
	
	In addition to model interpretability, we also visualize variable selection results of the ORL dataset as examples to demonstrate post hoc interpretability of TNVS. Under the premise of the transparent search, the results of TNVS are intuitionistic.
	The literature on facial landmark detection has highlighted some key reference points on the face that could help identity an individual, e.g., the Annotated Facial Landmarks in the Wild (AFLW) marked $21$ facial feature points \citep{kostingerAnnotatedFacialLandmarks2011}, including $12$ points on the eyes and eyebrows, $3$ on the mouth, $3$ on the nose, $3$ on each ear, and $1$ on the chin. Checking whether our selected pixels cover most if not all of these key reference points, we can observe qualitatively whether the selection results provide insight to capture the key predictors of these observations.
	
\begin{figure}[H]
		\centering 
		\includegraphics[width=0.85\linewidth]{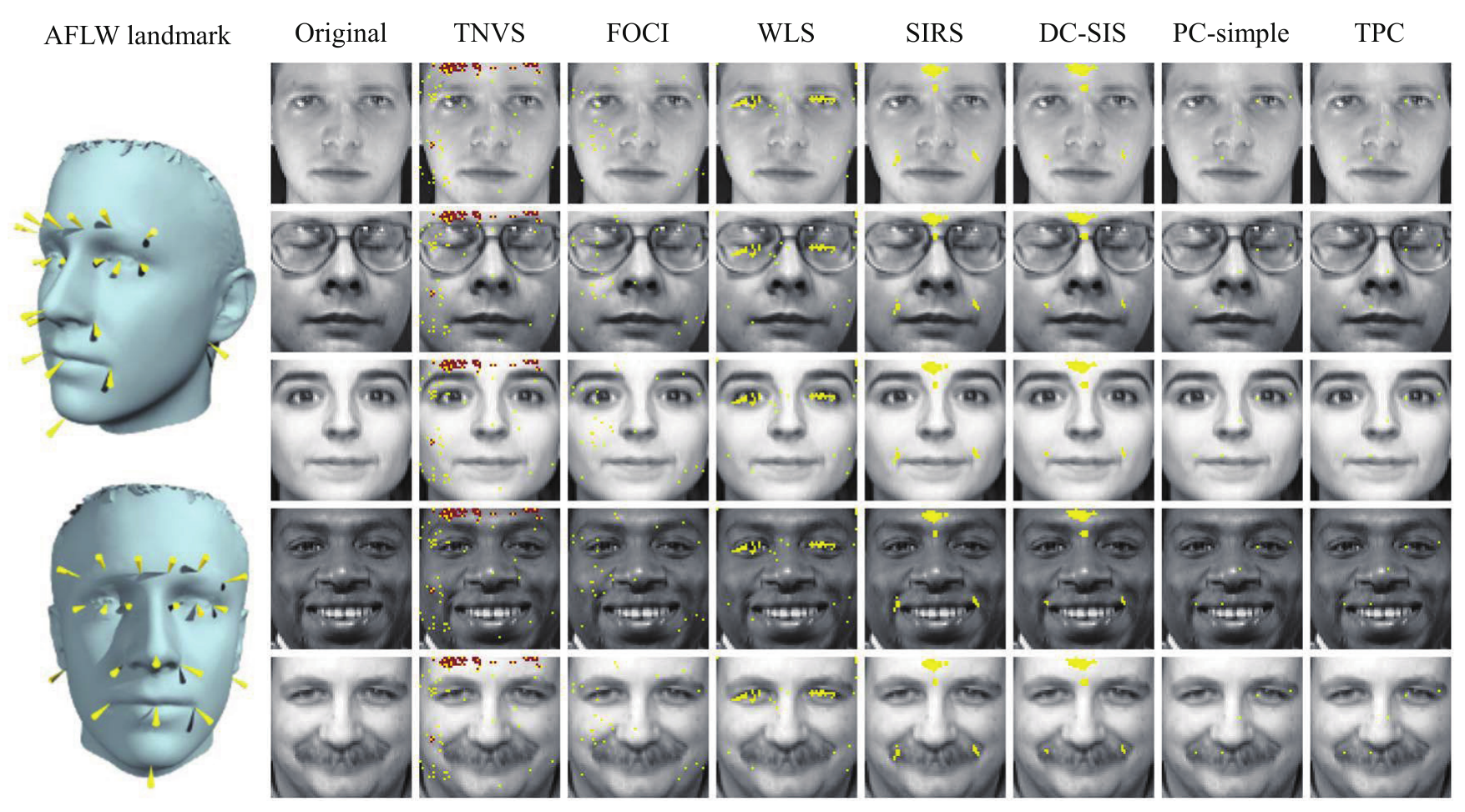}
		\caption{Variable selection results of TNVS and baselines are visualized on the original images of ORL. For TNVS, the relevant pixels being selected are marked in yellow, the redundant pixels being removed are marked in red, and the conditionally independent pixels being removed are unmarked on the original images. For others, the selected pixels are marked in yellow, and the deleted ones are unmarked.}
		\label{fig:ORL_feature_selection}
	\end{figure}

	One of the variable selection results is marked on the original images of ORL. For TNVS, we color the relevant pixels in yellow, the redundant pixels in red, and the unmarked pixels are conditionally independent pixels, as shown in Figure \ref{fig:ORL_feature_selection}. For other methods, we color the selected pixels in yellow, and the unselected pixels are left unmarked. For ORL, TNVS selects $62$ relevant pixels, and removes $66$ redundant pixels, and $3\ 968$ conditionally independent pixels. FOCI selects $26$ pixels, WLS, DC-SIS, and SIRS select $62$ pixels, and PC-simple and TPC select $5$ pixels. Compared with the predictive results of baselines in Table \ref{tab:appendix_A1} of \nameref{appendix_1}, TNVS achieves comparable or even better performance with more concise subset of predictors. 
	
	As is shown in Table \ref{tab:appendix_A1} of \nameref{appendix_1}, TNVS and WLS are the best two methods on ORL dataset. Their selected pixels both cover the facial points of AFLW, which mainly lie in eye corners, eyebrow corners, tips of nose, corners of mouth, and chins. The accurate detection of key reference points intuitively explains the superior facial discrimination of TNVS and WLS.
	TNVS further reveals that the redundant pixels of ORL are concentrated mainly in the foreheads, indicating that pixels in these areas are linearly correlated with the relevant pixels. Most regions belong to the conditionally independent subset, indicating that only a very small number of pixels is adequate to explain the response to a certain extent. Some pixels around the eye corners are usually symmetric, and TNVS only select pixels on one side, since given these areas, the pixels in symmetric areas will be conditionally independent. FOCI selects less pixels than TNVS, and these two methods select similar regions of the faces, but FOCI cannot identify the redundant pixels. SIRS and DC-SIS select a large number of redundant pixels in the forehead, and they ignore some feature points in the eye corners, eyebrow corners, and tips of the nose. PC-simple and TPC omit many important facial feature points.
	Together with the results in Table \ref{tab:appendix_A1} of \nameref{appendix_1}, this example shows that the predictors selected by TNVS can precisely predict the response, and the reserved or removed predictors can be explicitly interpreted.

	\subsection{Robustness in tuning the parameters}
	
	We further demonstrate stability of TNVS as the three thresholds vary. These ranges should be tuned within certain ranges. If these thresholds are too large, the predictors in each subset will violate their definitions, i.e., the uninformative subset may include informative predictors, and the redundant subset may include predictors which are not linearly independent. If the method is only robust within small ranges, it will be easily influnced by noise in real data. Here, we take a large enough range for each threshold to show that the outperformance is persistently significant, and the proposed method is robust against hyperparameter variations.
	
	First, the uninformative threshold is tuned within range $\alpha_1\in \{0,~0.01,\cdots,0.05\}$, and nonlinear correlations are evaluated between the response and the uninformative predictors $T_n(\mathbf{Y}, \mathbf{X}_{\mathcal{A}_1})$, and the nonlinear partial correlations between the response and uninformative predictors given the selected predictors $T_n(\mathbf{Y}, \mathbf{X}_{\mathcal{A}_1}\mid \mathbf{X}_\mathcal{S})$ for the four real datasets. If $T_n(\mathbf{Y}, \mathbf{X}_{\mathcal{A}_1}\mid \mathbf{X}_\mathcal{S})$ is undefined, it can be regarded that $\mathbf{Y}$ is almost surely a measurable function of $\mathbf{X}_\mathcal{S}$, and we set $T_n(\mathbf{Y}, \mathbf{X}_{\mathcal{A}_1}\mid \mathbf{X}_\mathcal{S})=0$. For warpAR10P and ORL, the numbers of uninformative predictors are $0$. For arcene and isolet, the numbers of identified uninformative predictors, their correlations and partial correlations are shown in Table \ref{tab:nonzero_parameter}. The number of uninformative predictors in arcene is the same when $\alpha_1\in [0,~0.03]$, and that in isolet is the same when $\alpha_1\in [0,~0.05]$. The correlations and partial correlations between the response and the uninformative predictors are close to $0$ in both datasets, which illustrates that the uninformative predictors have no positive effect on predicting the response. Identifying the uninformative subsets does not heavily depend on the settings of threshold $\alpha_1$, which shows the robustness of TNVS at the prefiltering step.

	\begin{table}[H]
		\centering
		\small
		\caption{Results of arcene and isolet as the uninformative threshold $\alpha_1$ changes, including the mean and standard deviation (in parentheses) of the number of uninformative predictors $\lvert \mathcal{A}_1 \rvert$, the CODEC of the response and all the uninformative predictors $T_n(\mathbf{Y}, \mathbf{X}_{\mathcal{A}_1})$, and the CODEC of the response and all the uninformative predictors given the selected subset $T_n(\mathbf{Y}, \mathbf{X}_{\mathcal{A}_1}\mid \mathbf{X}_\mathcal{S})$. }
		\begin{tabular}{lccccccc}
			\toprule
			Dataset & Indicator & $\alpha_1=0$ & $\alpha_1=0.01$ & $\alpha_1=0.02$ & $\alpha_1=0.03$ & $\alpha_1=0.04$ & $\alpha_1=0.05$ \\ 
			\midrule
			\multirow{3}{*}{arcene} & $\lvert \mathcal{A}_1 \rvert$ & $\underset{(5.95)}{45.72}$ & $\underset{(5.95)}{45.72}$ & $\underset{(5.95)}{45.72}$ & $\underset{(5.95)}{45.72}$ & $\underset{(7.34)}{111.25}$ & $\underset{(7.34)}{111.25}$ \\
			& $T_n(\mathbf{Y}, \mathbf{X}_{\mathcal{A}_1})$ & $\underset{(0.118)}{0.006}$ & $\underset{(0.088)}{-0.009}$ & $\underset{(0.111)}{-0.010}$ & $\underset{(0.107)}{0.003}$ & $\underset{(0.105)}{-0.037}$ & $\underset{(0.097)}{-0.025}$ \\
			& $T_n(\mathbf{Y}, \mathbf{X}_{\mathcal{A}_1}\mid \mathbf{X}_\mathcal{S})$ & $\underset{(0.058)}{0.006}$ & $\underset{(0.066)}{0.012}$ & $\underset{(0.029)}{0.001}$ & $\underset{(0.207)}{-0.022}$ & $\underset{(2.888)}{-1.899}$ & $\underset{(2.894)}{-1.901}$ \\
			\midrule
			\multirow{3}{*}{isolet} & $\lvert \mathcal{A}_1 \rvert$ & $\underset{(0.00)}{2.00}$ & $\underset{(0.00)}{2.00}$ & $\underset{(0.00)}{2.00}$ & $\underset{(0.00)}{2.00}$ & $\underset{(0.00)}{2.00}$ & $\underset{(0.00)}{2.00}$ \\
			& $T_n(\mathbf{Y}, \mathbf{X}_{\mathcal{A}_1})$ & $\underset{(0.064)}{-0.037}$ & $\underset{(0.067)}{-0.034}$ & $\underset{(0.061)}{-0.027}$ & $\underset{(0.059)}{-0.035}$ & $\underset{(0.062)}{-0.025}$ & $\underset{(0.065)}{-0.036}$ \\
			& $T_n(\mathbf{Y}, \mathbf{X}_{\mathcal{A}_1}\mid \mathbf{X}_\mathcal{S})$ & $\underset{(4.564)}{-2.822}$ & $\underset{(4.564)}{-2.822}$ & $\underset{(4.564)}{-2.822}$ & $\underset{(4.564)}{-2.822}$ & $\underset{(4.564)}{-2.822}$ & $\underset{(4.564)}{-2.822}$ \\
			\bottomrule
		\end{tabular}
		\label{tab:nonzero_parameter}
	\end{table}
	
\begin{figure}[H]
	\centering
	\includegraphics[width=0.8\linewidth]{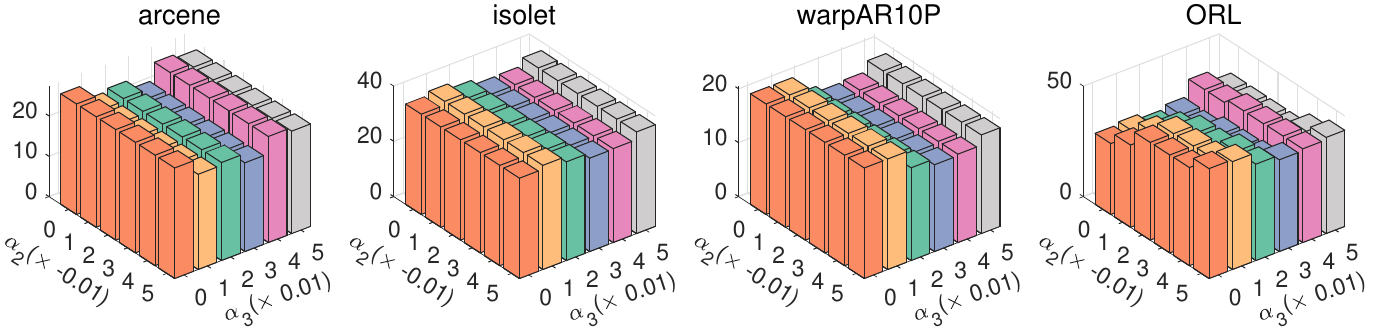}
	\caption{The average number of predictors retained by TNVS under different parameter settings.}
	\label{fig:parameter_adj_red}
	\end{figure}

	We next tune the remaining two parameters by taking the value of the relevant threshold $\alpha_2\in\{-0.05,~-0.04,\cdots,0\}$ and the value of the redundant threshold $\alpha_3\in\{0,~0.01,\cdots,0.05\}$. For each combination of $\alpha_2$ and $\alpha_3$, a 10-fold cross-validation is performed on the four real datasets to demonstrate the stability of TNVS as these two parameters vary. Fig. \ref{fig:parameter_adj_red} shows the average cardinality of the selected subsets obtained by TNVS under different values of $\alpha_2$ and $\alpha_3$. The number of selected predictors fluctuates slightly in response to the change in $\alpha_2$ and $\alpha_3$, and there is no obvious and unified trend in different datasets. Fig. \ref{fig:parameter_adj_pred} of \nameref{appendix_2} further describes the prediction accuracy using the above selected predictors. As expected, the changes in both thresholds have little influence on the predictive results. In summary, the proposed TNVS is insensitive to changes in the thresholds within certain ranges.
	
	\section{Conclusions and future work} \label{Conclusion}
	In this paper, a Transparent and Nonlinear Variable Selection (TNVS) method was proposed for high-dimensional data. 
	Transparent information decoupling was achieved with a three-step heuristic search, where the predictors which were relevant to the response were selected, and uninformative, collinear, and conditionally independent predictors were deleted. Introducing a recently proposed nonlinear partial correlation, TNVS was able to identify very complex nonlinear functional dependencies, including not only monotonic and additive functional dependencies between the response and predictors, but also nonmonotonic or oscillatory functional dependence and interactions among predictors. 
	According to scores based on information entropy, Gram-Schmidt orthogonalization, and nonlinear partial correlation, the unimportant predictors were classified into three unimportant subsets. The clear selecting and deleting process enhanced the effectiveness and model interpretability of the proposed method for variable selection.
	
	We should note that limitations exist in the proposed method, and extensions could be developed. 
	First, the three thresholds were designed empirically, although the selection kept stable within certain ranges of these thresholds. In the future, more advanced measures can be introduced to improve the empirical thresholds by using techniques such as nonparametric statistical tests \citep{shiAzadkiaChatterjeeConditionalDependence2021}. 
	In addition, though the proposed method made progress in identifying the interacted predictors, it had difficulty in identifying the marginal independent but jointly correlated predictors. In the future, domain knowledge can be considered in the initial stage to better support the detection of such hard-to-identified interactions \citep{wuDomainKnowledgeenhancedVariable2022}.
	Last but not least, although the three-step heuristic search adopted in the proposed method was more efficient than other stepwise methods, it could be time-consuming when a large number of predictors were relevant to the response. In the future, more advanced algorithms such as genetic algorithm \citep{saibeneGeneticAlgorithmfor2023} and other metaheuristic algorithms \citep{alcarazSupportVectorMachine2022, pramanikBreastCancerDetection2023} can be designed to improve the efficiency of TNVS in a less sparse high-dimensional scenario.

	\section*{Acknowledgments} \label{Acknowledgments}
	The work was supported by grants from the National Natural Science Foundation of China (Grant Nos. 72021001 and 71871006).
	
	\section*{CRediT authorship contribution statement}
	\textbf{Keyao Wang}: Conceptualization, Methodology, Software, Writing - Original draft preparation.
	\textbf{Huiwen Wang}: Conceptualization, Resources, Supervision, Writing - Reviewing and Editing.
	\textbf{Jichang Zhao}: Validation, Writing - Reviewing and Editing. 
	\textbf{Lihong Wang}: Supervision, Resources, Writing - Reviewing and Editing.

	\section*{Declaration of Competing Interest}
	The authors declare that they have no known competing financial interests or personal relationships that could have appeared
	to influence the work reported in this paper.

\appendix 
	\section*{Appendix A} \label{appendix_0}
	\setcounter{table}{0}
	\setcounter{figure}{0}
	\renewcommand{\thetable}{A\arabic{table}}
	\renewcommand{\thefigure}{A\arabic{figure}}
	\renewcommand{\thesubsection}{A\arabic{subsection}}
A simple example is provided to further demonstrate the measures and the three-step heuristic search of the proposed method for variable selection.
\subsection{Relevant, uninformative, multicollinear, and conditionally independent predictors} \label{subsec:A1}
	We generate a regression example with nonlinear relevance, uninformative predictors, multicollinearity, and conditionally independent predictors. The predictors and response are defined as follows:
	
	\begin{enumerate}[(1)]
		\item $X_1$, $X_2$, and $X_3$ are mutually independent predictors all sampled from the standard normal distribution $\mathcal{N}(0,~1)$. 
		\item $X_4 = X_1 + X_2$, and $X_5 = X_1 + X_3$. 
		\item only $0.1\%$ observations of $X_6$ are nonzero, and these nonzero observations are randomly generated from $\mathcal{N}(0,~ 0.1^2)$. 
		\item $Y = X_1\cdot X_2$.
	\end{enumerate}
	
\begin{figure}[htbp]
	\centering 
	\subfloat[Histogram of $X_6$. \label{fig:A1a}]{
		\includegraphics[width=0.32\linewidth]{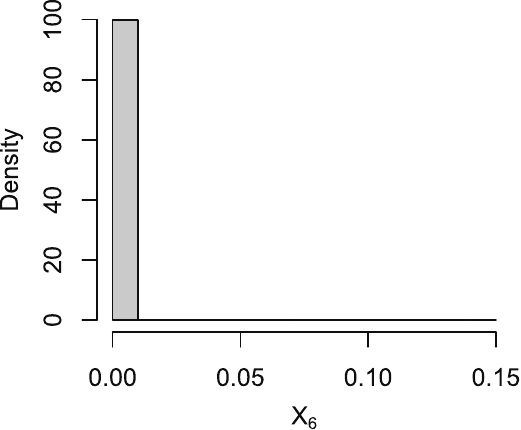}}
	\subfloat[Dependency of $Y$ and $X_2$ given $X_1$. \label{fig:A1b}]{
		\includegraphics[width=0.32\linewidth]{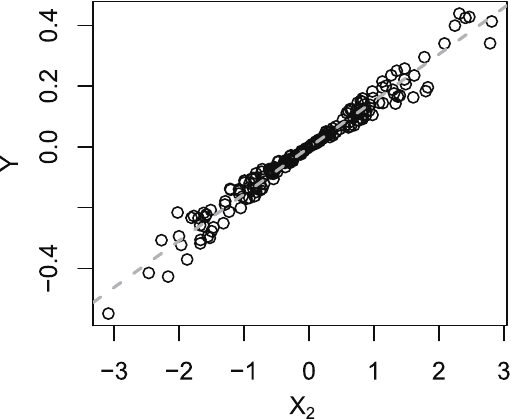}}
	\hspace*{\fill}
	\subfloat[Dependency of $Y$ and $X_5$ given $X_1$.\label{fig:A1c}]{
		\includegraphics[width=0.32\linewidth]{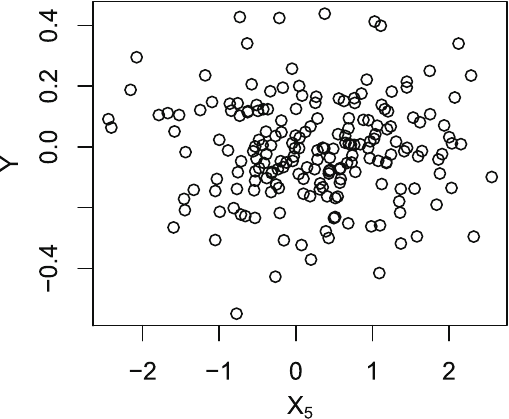}} 
	\caption{An example to illustrate features of the uninformative predictor $X_6$, relevant predictor $X_2$, and conditionally independent predictor $X_5$.}
	\label{fig:example}
	\end{figure}

	In this example, $X_6$ is an uninformative predictor, and its histogram is shown in Fig. \ref{fig:A1a}. $Y$ is nonlinearly correlated with $X_1$ and $X_2$, or any combination of their collinear predictors, i.e., $X_1$ and $X_4$, $X_2$ and $X_4$. There are interactions between $X_1$ and $X_2$, and given $X_1$, $Y$ is functionally dependent on $X_2$. Thus, if $X_1$ and $X_2$ are identified as relevant predictors, $X_4$ is the redundant predictor which is collinear with $X_1$ and $X_2$. Given the relevant predictors, $X_3$ and $X_5$ are conditionally independent predictors. The dependency between $Y$ and $X_2$ given $X_1\in[0.1,~0.2]$ is shown in Fig. \ref{fig:A1b}, and that between $Y$ and $X_5$ given $X_1\in[0.1,~0.2]$ is shown in Fig. \ref{fig:A1c}.
	
\subsection{The procedure of the proposed variable selection method} \label{subsec:A2}
	The heuristic search of TNVS includes prefiltering, forward selection, and batch deletion steps. According to the flowchart in Fig. \ref{fig:framework_of_TNVS}, the procedure of the proposed method is as follows. In the prefiltering step, uninformative predictors, i.e., $X_6$ in the example is identified and removed according to UniS. In the first iteration of forward selection, the predictor with largest RelS will be selected. Assume $X_1$ is the selected relevant predictor. Then, a deletion step is performed to remove redundant predictors that are collinear with $X_1$. Here, none of the predictors are identified. In the second iteration, assume $X_2$ is selected as the relevant predictor, and then $X_4$ will be identified as redundant predictor since it is the linear combination of $X_1$ and $X_2$. After two iterations, RelSs of the rest predictors $X_3$ and $X_5$ are both less than $0$. The search will stop, and the rest two predictors are conditionally independent. The inputs are separated into the uninformative subset $\{X_6\}$, the relevant subset $\{X_1,~X_2\}$, the redundant subset $\{X_4\}$, and the conditionally independent subset $\{X_3,~X_5\}$.

\section*{Appendix B} \label{appendix_1}
\setcounter{table}{0}
\setcounter{figure}{0}
\renewcommand{\thetable}{B\arabic{table}}
\renewcommand{\thefigure}{B\arabic{figure}}

\begin{center}
	\small
	\begin{longtable}{cccccccccc}
		\caption{The mean and standard deviation (in parentheses) of the predictive results in real datasets. The best results are presented in bold, and the second-best results are underlined.} \\
\toprule
Dataset & Model & Indicator & TNVS & FOCI & WLS & SIRS & DC-SIS & PC-simple & TPC \\ 
\midrule
& 
\multirow{7}{*}{SVM} & ACC & $\underset{(0.1092)}{0.6790}$ & $\underset{(0.1092)}{0.6880}$ & $\underset{(0.1161)}{0.6200}$ & $\underset{(0.0980)}{\mathbf{0.6955}}$ & $\underset{(0.1029)}{\underline{0.6920}}$ & $\underset{(0.0919)}{0.6575}$ & $\underset{(0.1014)}{0.6515}$ \\ 
&  & recall & $\underset{(0.1099)}{0.6820}$ & $\underset{(0.1099)}{0.6924}$ & $\underset{(0.1222)}{0.6007}$ & $\underset{(0.0994)}{\mathbf{0.6980}}$ & $\underset{(0.1038)}{\underline{0.6928}}$ & $\underset{(0.0925)}{0.6512}$ & $\underset{(0.0990)}{0.6649}$ \\ 
&  & $F_1$ & $\underset{(0.1146)}{0.6704}$ & $\underset{(0.1146)}{0.6804}$ & $\underset{(0.1418)}{0.5805}$ & $\underset{(0.1007)}{\mathbf{0.6901}}$ & $\underset{(0.1056)}{\underline{0.6854}}$ & $\underset{(0.0946)}{0.6458}$ & $\underset{(0.1035)}{0.6470}$ \\ 
&  & Kappa & $\underset{(0.2151)}{0.3579}$ & $\underset{(0.2151)}{0.3778}$ & $\underset{(0.2474)}{0.2034}$ & $\underset{(0.1970)}{\mathbf{0.3898}}$ & $\underset{(0.2050)}{\underline{0.3810}}$ & $\underset{(0.1858)}{0.3025}$ & $\underset{(0.1952)}{0.3191}$ \\ 
\cline{2-10}
\specialrule{0em}{1pt}{1pt}
& \multirow{7}[1]{*}{RF} & ACC & $\underset{(0.0973)}{\mathbf{0.7415}}$ & $\underset{(0.0973)}{\underline{0.7350}}$ & $\underset{(0.1310)}{0.6885}$ & $\underset{(0.0990)}{0.6705}$ & $\underset{(0.0990)}{0.6790}$ & $\underset{(0.0957)}{0.6910}$ & $\underset{(0.0939)}{0.5870}$ \\ 
&  & recall & $\underset{(0.1019)}{\mathbf{0.7383}}$ & $\underset{(0.1019)}{\underline{0.7290}}$ & $\underset{(0.1327)}{0.6805}$ & $\underset{(0.1029)}{0.6658}$ & $\underset{(0.1009)}{0.6743}$ & $\underset{(0.0986)}{0.6845}$ & $\underset{(0.0923)}{0.6059}$ \\ 
&  & $F_1$ & $\underset{(0.1041)}{\mathbf{0.7348}}$ & $\underset{(0.1041)}{\underline{0.7254}}$ & $\underset{(0.1425)}{0.6730}$ & $\underset{(0.1042)}{0.6609}$ & $\underset{(0.1025)}{0.6703}$ & $\underset{(0.1014)}{0.6807}$ & $\underset{(0.0990)}{0.5768}$ \\ 
&  & Kappa & $\underset{(0.2033)}{\mathbf{0.4754}}$ & $\underset{(0.2033)}{\underline{0.4579}}$ & $\underset{(0.2673)}{0.3623}$ & $\underset{(0.2049)}{0.3301}$ & $\underset{(0.2018)}{0.3476}$ & $\underset{(0.1972)}{0.3690}$ & $\underset{(0.1777)}{0.2020}$ \\ 
\cline{2-10}
\specialrule{0em}{1pt}{1pt}
\multirow{7}{*}{arcene}  & \multirow{7}{*}{DT} & ACC & $\underset{(0.1093)}{\underline{0.6945}}$ & $\underset{(0.1093)}{\mathbf{0.7050}}$ & $\underset{(0.1176)}{0.6420}$ & $\underset{(0.1100)}{0.6495}$ & $\underset{(0.0969)}{0.6555}$ & $\underset{(0.0960)}{0.6375}$ & $\underset{(0.0938)}{0.5875}$ \\ 
&  & recall & $\underset{(0.1123)}{\underline{0.6911}}$ & $\underset{(0.1123)}{\mathbf{0.7000}}$ & $\underset{(0.1200)}{0.6347}$ & $\underset{(0.1105)}{0.6444}$ & $\underset{(0.0979)}{0.6485}$ & $\underset{(0.0991)}{0.6344}$ & $\underset{(0.0932)}{0.6038}$ \\ 
&  & $F_1$ & $\underset{(0.1156)}{\underline{0.6841}}$ & $\underset{(0.1156)}{\mathbf{0.6946}}$ & $\underset{(0.1246)}{0.6275}$ & $\underset{(0.1121)}{0.6412}$ & $\underset{(0.0999)}{0.6453}$ & $\underset{(0.1007)}{0.6282}$ & $\underset{(0.0980)}{0.5785}$ \\ 
&  & Kappa & $\underset{(0.2246)}{\underline{0.3808}}$ & $\underset{(0.2246)}{\mathbf{0.3993}}$ & $\underset{(0.2403)}{0.2692}$ & $\underset{(0.2209)}{0.2888}$ & $\underset{(0.1976)}{0.2977}$ & $\underset{(0.1962)}{0.2662}$ & $\underset{(0.1804)}{0.1988}$ \\ 
\cline{2-10}
\specialrule{0em}{1pt}{1pt}
& \multirow{7}{*}{lightGBM} & ACC & $\underset{(0.1027)}{\mathbf{0.7270}}$ & $\underset{(0.1027)}{\underline{0.7185}}$ & $\underset{(0.1183)}{0.6915}$ & $\underset{(0.1027)}{0.6695}$ & $\underset{(0.0941)}{0.6875}$ & $\underset{(0.0982)}{0.6800}$ & $\underset{(0.1145)}{0.6345}$ \\ 
&  & recall & $\underset{(0.1049)}{\mathbf{0.7249}}$ & $\underset{(0.1049)}{\underline{0.7155}}$ & $\underset{(0.1227)}{0.6821}$ & $\underset{(0.1069)}{0.6651}$ & $\underset{(0.0958)}{0.6851}$ & $\underset{(0.1013)}{0.6752}$ & $\underset{(0.1117)}{0.6491}$ \\ 
&  & $F_1$ & $\underset{(0.1066)}{\mathbf{0.7197}}$ & $\underset{(0.1066)}{\underline{0.7108}}$ & $\underset{(0.1381)}{0.6716}$ & $\underset{(0.1081)}{0.6606}$ & $\underset{(0.0964)}{0.6808}$ & $\underset{(0.1036)}{0.6697}$ & $\underset{(0.1171)}{0.6293}$ \\ 
&  & Kappa & $\underset{(0.2076)}{\mathbf{0.4476}}$ & $\underset{(0.2076)}{\underline{0.4291}}$ & $\underset{(0.2463)}{0.3651}$ & $\underset{(0.2128)}{0.3281}$ & $\underset{(0.1913)}{0.3676}$ & $\underset{(0.2016)}{0.3490}$ & $\underset{(0.2187)}{0.2884}$ \\ 
\cline{2-10}
\specialrule{0em}{1pt}{1pt}
& \multirow{7}{*}{MLP} & ACC & $\underset{(0.1122)}{0.7040}$ & $\underset{(0.1122)}{\mathbf{0.7210}}$ & $\underset{(0.1332)}{0.6395}$ & $\underset{(0.1114)}{\underline{0.7155}}$ & $\underset{(0.1093)}{0.6910}$ & $\underset{(0.0968)}{0.6505}$ & $\underset{(0.1005)}{0.6570}$ \\ 
&  & recall & $\underset{(0.1130)}{0.7022}$ & $\underset{(0.1130)}{\mathbf{0.7202}}$ & $\underset{(0.1335)}{0.6329}$ & $\underset{(0.1131)}{\underline{0.7110}}$ & $\underset{(0.1132)}{0.6867}$ & $\underset{(0.0979)}{0.6467}$ & $\underset{(0.0990)}{0.6692}$ \\ 
&  & $F_1$ & $\underset{(0.1147)}{0.6967}$ & $\underset{(0.1147)}{\mathbf{0.7150}}$ & $\underset{(0.1407)}{0.6248}$ & $\underset{(0.1153)}{\underline{0.7070}}$ & $\underset{(0.1147)}{0.6811}$ & $\underset{(0.0999)}{0.6401}$ & $\underset{(0.1020)}{0.6529}$ \\ 
&  & Kappa & $\underset{(0.2237)}{0.4007}$ & $\underset{(0.2237)}{\mathbf{0.4373}}$ & $\underset{(0.2691)}{0.2661}$ & $\underset{(0.2258)}{\underline{0.4216}}$ & $\underset{(0.2241)}{0.3712}$ & $\underset{(0.1947)}{0.2921}$ & $\underset{(0.1953)}{0.3278}$ \\ 
\midrule
& \multirow{7}{*}{SVM} & ACC & $\underset{(0.0966)}{\underline{0.8723}}$ & $\underset{(0.0966)}{0.8203}$ & $\underset{(0.0577)}{\mathbf{0.8860}}$ & $\underset{(0.0705)}{0.4877}$ & $\underset{(0.0729)}{0.4973}$ & $\underset{(0.0749)}{0.4630}$ & $\underset{(0.0810)}{0.4143}$ \\ 
&  & recall & $\underset{(0.0966)}{\underline{0.8723}}$ & $\underset{(0.0966)}{0.8203}$ & $\underset{(0.0577)}{\mathbf{0.8860}}$ & $\underset{(0.0705)}{0.4877}$ & $\underset{(0.0729)}{0.4973}$ & $\underset{(0.0749)}{0.4630}$ & $\underset{(0.0810)}{0.4143}$ \\ 
&  & $F_1$ & $\underset{(0.1013)}{\underline{0.8687}}$ & $\underset{(0.1013)}{0.8157}$ & $\underset{(0.0583)}{\mathbf{0.8846}}$ & $\underset{(0.0676)}{0.4688}$ & $\underset{(0.0677)}{0.4766}$ & $\underset{(0.0782)}{0.4454}$ & $\underset{(0.0797)}{0.3990}$ \\ 
&  & Kappa & $\underset{(0.1159)}{\underline{0.8468}}$ & $\underset{(0.1159)}{0.7844}$ & $\underset{(0.0692)}{\mathbf{0.8632}}$ & $\underset{(0.0846)}{0.3852}$ & $\underset{(0.0875)}{0.3968}$ & $\underset{(0.0899)}{0.3556}$ & $\underset{(0.0972)}{0.2972}$ \\ 
\cline{2-10}
\specialrule{0em}{1pt}{1pt}
& \multirow{7}{*}{RF} & ACC & $\underset{(0.0965)}{\mathbf{0.8733}}$ & $\underset{(0.0965)}{0.8297}$ & $\underset{(0.0613)}{\underline{0.8727}}$ & $\underset{(0.0734)}{0.4890}$ & $\underset{(0.0698)}{0.4963}$ & $\underset{(0.0737)}{0.4347}$ & $\underset{(0.0775)}{0.3720}$ \\ 
&  & recall & $\underset{(0.0965)}{\mathbf{0.8733}}$ & $\underset{(0.0965)}{0.8297}$ & $\underset{(0.0613)}{\underline{0.8727}}$ & $\underset{(0.0734)}{0.4890}$ & $\underset{(0.0698)}{0.4963}$ & $\underset{(0.0737)}{0.4347}$ & $\underset{(0.0775)}{0.3720}$ \\ 
&  & $F_1$ & $\underset{(0.0981)}{\underline{0.8709}}$ & $\underset{(0.0981)}{0.8263}$ & $\underset{(0.0625)}{\mathbf{0.8712}}$ & $\underset{(0.0756)}{0.4736}$ & $\underset{(0.0713)}{0.4782}$ & $\underset{(0.0748)}{0.4216}$ & $\underset{(0.0759)}{0.3594}$ \\ 
&  & Kappa & $\underset{(0.1158)}{\mathbf{0.8480}}$ & $\underset{(0.1158)}{0.7956}$ & $\underset{(0.0736)}{\underline{0.8472}}$ & $\underset{(0.0881)}{0.3868}$ & $\underset{(0.0837)}{0.3956}$ & $\underset{(0.0884)}{0.3216}$ & $\underset{(0.0930)}{0.2464}$ \\ 
\cline{2-10}
\specialrule{0em}{1pt}{1pt}
\multirow{7}{*}{isolet} & \multirow{7}{*}{DT} & ACC & $\underset{(0.1081)}{\mathbf{0.8113}}$ & $\underset{(0.1081)}{0.7857}$ & $\underset{(0.0829)}{\underline{0.7893}}$ & $\underset{(0.0712)}{0.4270}$ & $\underset{(0.0813)}{0.4353}$ & $\underset{(0.0785)}{0.3960}$ & $\underset{(0.0772)}{0.3547}$ \\ 
&  & recall & $\underset{(0.1081)}{\mathbf{0.8113}}$ & $\underset{(0.1081)}{0.7857}$ & $\underset{(0.0829)}{\underline{0.7893}}$ & $\underset{(0.0712)}{0.4270}$ & $\underset{(0.0813)}{0.4353}$ & $\underset{(0.0785)}{0.3960}$ & $\underset{(0.0772)}{0.3547}$ \\ 
&  & $F_1$ & $\underset{(0.1105)}{\mathbf{0.8076}}$ & $\underset{(0.1105)}{0.7815}$ & $\underset{(0.0856)}{\underline{0.7835}}$ & $\underset{(0.0705)}{0.4194}$ & $\underset{(0.0773)}{0.4297}$ & $\underset{(0.0770)}{0.3875}$ & $\underset{(0.0729)}{0.3455}$ \\ 
&  & Kappa & $\underset{(0.1297)}{\mathbf{0.7736}}$ & $\underset{(0.1297)}{0.7428}$ & $\underset{(0.0994)}{\underline{0.7472}}$ & $\underset{(0.0854)}{0.3124}$ & $\underset{(0.0976)}{0.3224}$ & $\underset{(0.0942)}{0.2752}$ & $\underset{(0.0926)}{0.2256}$ \\ 
\cline{2-10}
\specialrule{0em}{1pt}{1pt}
& \multirow{7}{*}{lightGBM} & ACC & $\underset{(0.1108)}{\mathbf{0.8807}}$ & $\underset{(0.1108)}{0.8197}$ & $\underset{(0.0645)}{\underline{0.8787}}$ & $\underset{(0.0789)}{0.4953}$ & $\underset{(0.0749)}{0.5133}$ & $\underset{(0.0731)}{0.4130}$ & $\underset{(0.0646)}{0.3647}$ \\ 
&  & recall & $\underset{(0.1108)}{\mathbf{0.8807}}$ & $\underset{(0.1108)}{0.8197}$ & $\underset{(0.0645)}{\underline{0.8787}}$ & $\underset{(0.0789)}{0.4953}$ & $\underset{(0.0749)}{0.5133}$ & $\underset{(0.0731)}{0.4130}$ & $\underset{(0.0646)}{0.3647}$ \\ 
&  & $F_1$ & $\underset{(0.1101)}{\mathbf{0.8784}}$ & $\underset{(0.1101)}{0.8188}$ & $\underset{(0.0648)}{\underline{0.8777}}$ & $\underset{(0.0794)}{0.4845}$ & $\underset{(0.0765)}{0.5012}$ & $\underset{(0.0713)}{0.4031}$ & $\underset{(0.0644)}{0.3563}$ \\ 
&  & Kappa & $\underset{(0.1330)}{\mathbf{0.8568}}$ & $\underset{(0.1330)}{0.7836}$ & $\underset{(0.0774)}{\underline{0.8544}}$ & $\underset{(0.0946)}{0.3944}$ & $\underset{(0.0899)}{0.4160}$ & $\underset{(0.0877)}{0.2956}$ & $\underset{(0.0775)}{0.2376}$ \\ 
\cline{2-10}
\specialrule{0em}{1pt}{1pt}
& \multirow{7}{*}{MLP} & ACC & $\underset{(0.0930)}{\underline{0.8813}}$ & $\underset{(0.0930)}{0.8237}$ & $\underset{(0.0556)}{\mathbf{0.8980}}$ & $\underset{(0.0808)}{0.4730}$ & $\underset{(0.0795)}{0.4930}$ & $\underset{(0.0752)}{0.4227}$ & $\underset{(0.0763)}{0.3883}$ \\ 
&  & recall & $\underset{(0.0930)}{\underline{0.8813}}$ & $\underset{(0.0930)}{0.8237}$ & $\underset{(0.0556)}{\mathbf{0.8980}}$ & $\underset{(0.0808)}{0.4730}$ & $\underset{(0.0795)}{0.4930}$ & $\underset{(0.0752)}{0.4227}$ & $\underset{(0.0763)}{0.3883}$ \\ 
&  & $F_1$ & $\underset{(0.0965)}{\underline{0.8791}}$ & $\underset{(0.0965)}{0.8206}$ & $\underset{(0.0566)}{\mathbf{0.8967}}$ & $\underset{(0.0771)}{0.4664}$ & $\underset{(0.0766)}{0.4873}$ & $\underset{(0.0762)}{0.4150}$ & $\underset{(0.0764)}{0.3786}$ \\ 
&  & Kappa & $\underset{(0.1116)}{\underline{0.8576}}$ & $\underset{(0.1116)}{0.7884}$ & $\underset{(0.0667)}{\mathbf{0.8776}}$ & $\underset{(0.0969)}{0.3676}$ & $\underset{(0.0954)}{0.3916}$ & $\underset{(0.0902)}{0.3072}$ & $\underset{(0.0916)}{0.2660}$ \\ 
\hline
\specialrule{0em}{2pt}{1pt}
& \multirow{7}{*}{SVM} & ACC & $\underset{(0.1371)}{\mathbf{0.7015}}$ & $\underset{(0.1371)}{\underline{0.6545}}$ & $\underset{(0.1477)}{0.5073}$ & $\underset{(0.1317)}{0.3833}$ & $\underset{(0.1211)}{0.3845}$ & $\underset{(0.1356)}{0.3423}$ & $\underset{(0.1389)}{0.3495}$ \\ 
&  & recall & $\underset{(0.1371)}{\mathbf{0.7015}}$ & $\underset{(0.1371)}{\underline{0.6545}}$ & $\underset{(0.1477)}{0.5073}$ & $\underset{(0.1317)}{0.3833}$ & $\underset{(0.1211)}{0.3845}$ & $\underset{(0.1356)}{0.3423}$ & $\underset{(0.1389)}{0.3495}$ \\ 
&  & $F_1$ & $\underset{(0.1519)}{\mathbf{0.6434}}$ & $\underset{(0.1519)}{\underline{0.5902}}$ & $\underset{(0.1464)}{0.4524}$ & $\underset{(0.1334)}{0.3108}$ & $\underset{(0.1190)}{0.3075}$ & $\underset{(0.1226)}{0.2778}$ & $\underset{(0.1218)}{0.2840}$ \\ 
&  & Kappa & $\underset{(0.1523)}{\mathbf{0.6683}}$ & $\underset{(0.1523)}{\underline{0.6161}}$ & $\underset{(0.1641)}{0.4525}$ & $\underset{(0.1463)}{0.3147}$ & $\underset{(0.1345)}{0.3161}$ & $\underset{(0.1507)}{0.2692}$ & $\underset{(0.1543)}{0.2772}$ \\ 
\cline{2-10}
\specialrule{0em}{1pt}{1pt}
& \multirow{7}{*}{RF} & ACC & $\underset{(0.1282)}{\mathbf{0.7920}}$ & $\underset{(0.1282)}{\underline{0.7515}}$ & $\underset{(0.1427)}{0.6350}$ & $\underset{(0.1479)}{0.6505}$ & $\underset{(0.1494)}{0.6523}$ & $\underset{(0.1396)}{0.3453}$ & $\underset{(0.1443)}{0.3680}$ \\ 
&  & recall & $\underset{(0.1282)}{\mathbf{0.7920}}$ & $\underset{(0.1282)}{\underline{0.7515}}$ & $\underset{(0.1427)}{0.6350}$ & $\underset{(0.1479)}{0.6505}$ & $\underset{(0.1494)}{0.6523}$ & $\underset{(0.1396)}{0.3453}$ & $\underset{(0.1443)}{0.3680}$ \\ 
&  & $F_1$ & $\underset{(0.1487)}{\mathbf{0.7434}}$ & $\underset{(0.1487)}{\underline{0.6968}}$ & $\underset{(0.1510)}{0.5727}$ & $\underset{(0.1562)}{0.5853}$ & $\underset{(0.1546)}{0.5869}$ & $\underset{(0.1283)}{0.2933}$ & $\underset{(0.1281)}{0.3088}$ \\ 
&  & Kappa & $\underset{(0.1424)}{\mathbf{0.7689}}$ & $\underset{(0.1424)}{\underline{0.7239}}$ & $\underset{(0.1585)}{0.5944}$ & $\underset{(0.1643)}{0.6117}$ & $\underset{(0.1660)}{0.6136}$ & $\underset{(0.1551)}{0.2725}$ & $\underset{(0.1603)}{0.2978}$ \\ 
\cline{2-10}
\specialrule{0em}{1pt}{1pt}
\multirow{7}{*}{warpAR10P} & \multirow{7}{*}{DT} & ACC & $\underset{(0.1420)}{\mathbf{0.6863}}$ & $\underset{(0.1420)}{\underline{0.6173}}$ & $\underset{(0.1422)}{0.5225}$ & $\underset{(0.1441)}{0.4453}$ & $\underset{(0.1421)}{0.4520}$ & $\underset{(0.1358)}{0.2955}$ & $\underset{(0.1309)}{0.3085}$ \\ 
&  & recall & $\underset{(0.1420)}{\mathbf{0.6863}}$ & $\underset{(0.142)}{\underline{0.6173}}$ & $\underset{(0.1422)}{0.5225}$ & $\underset{(0.1441)}{0.4453}$ & $\underset{(0.1421)}{0.4520}$ & $\underset{(0.1358)}{0.2955}$ & $\underset{(0.1309)}{0.3085}$ \\ 
&  & $F_1$ & $\underset{(0.1534)}{\mathbf{0.6309}}$ & $\underset{(0.1534)}{\underline{0.5527}}$ & $\underset{(0.1372)}{0.4582}$ & $\underset{(0.1346)}{0.3856}$ & $\underset{(0.1331)}{0.3864}$ & $\underset{(0.1181)}{0.2473}$ & $\underset{(0.1186)}{0.2576}$ \\ 
&  & Kappa & $\underset{(0.1578)}{\mathbf{0.6514}}$ & $\underset{(0.1578)}{\underline{0.5747}}$ & $\underset{(0.1580)}{0.4694}$ & $\underset{(0.1601)}{0.3836}$ & $\underset{(0.1579)}{0.3911}$ & $\underset{(0.1509)}{0.2172}$ & $\underset{(0.1454)}{0.2317}$ \\ 
\cline{2-10}
\specialrule{0em}{1pt}{1pt}
& \multirow{7}{*}{lightGBM} & ACC & $\underset{(0.1337)}{\mathbf{0.7735}}$ & $\underset{(0.1337)}{\underline{0.7115}}$ & $\underset{(0.1412)}{0.6273}$ & $\underset{(0.1452)}{0.5685}$ & $\underset{(0.1427)}{0.5655}$ & $\underset{(0.1379)}{0.2993}$ & $\underset{(0.1414)}{0.3363}$ \\ 
&  & recall & $\underset{(0.1337)}{\mathbf{0.7735}}$ & $\underset{(0.1337)}{\underline{0.7115}}$ & $\underset{(0.1412)}{0.6273}$ & $\underset{(0.1452)}{0.5685}$ & $\underset{(0.1427)}{0.5655}$ & $\underset{(0.1379)}{0.2993}$ & $\underset{(0.1414)}{0.3363}$ \\ 
&  & $F_1$ & $\underset{(0.1494)}{\mathbf{0.7266}}$ & $\underset{(0.1494)}{\underline{0.6551}}$ & $\underset{(0.1487)}{0.5647}$ & $\underset{(0.1468)}{0.5003}$ & $\underset{(0.1483)}{0.4988}$ & $\underset{(0.1223)}{0.2487}$ & $\underset{(0.1301)}{0.2798}$ \\ 
&  & Kappa & $\underset{(0.1486)}{\mathbf{0.7483}}$ & $\underset{(0.1486)}{\underline{0.6794}}$ & $\underset{(0.1568)}{0.5858}$ & $\underset{(0.1614)}{0.5206}$ & $\underset{(0.1586)}{0.5172}$ & $\underset{(0.1532)}{0.2214}$ & $\underset{(0.1571)}{0.2625}$ \\ 
\cline{2-10}
\specialrule{0em}{1pt}{1pt}
& \multirow{7}{*}{MLP} & ACC & $\underset{(0.1187)}{\mathbf{0.8248}}$ & $\underset{(0.1187)}{\underline{0.8003}}$ & $\underset{(0.1519)}{0.6550}$ & $\underset{(0.1429)}{0.6693}$ & $\underset{(0.1302)}{0.6713}$ & $\underset{(0.1257)}{0.3255}$ & $\underset{(0.1477)}{0.3603}$ \\ 
&  & recall & $\underset{(0.1187)}{\mathbf{0.8248}}$ & $\underset{(0.1187)}{\underline{0.8003}}$ & $\underset{(0.1519)}{0.6550}$ & $\underset{(0.1429)}{0.6693}$ & $\underset{(0.1302)}{0.6713}$ & $\underset{(0.1257)}{0.3255}$ & $\underset{(0.1477)}{0.3603}$ \\ 
&  & $F_1$ & $\underset{(0.1341)}{\mathbf{0.7826}}$ & $\underset{(0.1341)}{\underline{0.7591}}$ & $\underset{(0.1639)}{0.5934}$ & $\underset{(0.1506)}{0.6142}$ & $\underset{(0.1349)}{0.6157}$ & $\underset{(0.1163)}{0.2731}$ & $\underset{(0.1363)}{0.3007}$ \\ 
&  & Kappa & $\underset{(0.1319)}{\mathbf{0.8053}}$ & $\underset{(0.1319)}{\underline{0.7781}}$ & $\underset{(0.1688)}{0.6167}$ & $\underset{(0.1588)}{0.6325}$ & $\underset{(0.1447)}{0.6347}$ & $\underset{(0.1396)}{0.2506}$ & $\underset{(0.1641)}{0.2892}$ \\ 
\midrule
& \multirow{7}{*}{SVM} & ACC & $\underset{(0.1156)}{\underline{0.7905}}$ & $\underset{(0.1156)}{0.7428}$ & $\underset{(0.0346)}{\mathbf{0.9465}}$ & $\underset{(0.0657)}{0.6730}$ & $\underset{(0.0629)}{0.6638}$ & $\underset{(0.0829)}{0.4408}$ & $\underset{(0.0839)}{0.4433}$ \\ 
&  & recall & $\underset{(0.1156)}{\underline{0.7905}}$ & $\underset{(0.1156)}{0.7428}$ & $\underset{(0.0346)}{\mathbf{0.9465}}$ & $\underset{(0.0657)}{0.6730}$ & $\underset{(0.0629)}{0.6638}$ & $\underset{(0.0829)}{0.4408}$ & $\underset{(0.0839)}{0.4433}$ \\ 
&  & $F_1$ & $\underset{(0.1286)}{\underline{0.7486}}$ & $\underset{(0.1286)}{0.6931}$ & $\underset{(0.0438)}{\mathbf{0.9308}}$ & $\underset{(0.0705)}{0.6117}$ & $\underset{(0.0662)}{0.6020}$ & $\underset{(0.0828)}{0.3713}$ & $\underset{(0.0845)}{0.3731}$ \\ 
&  & Kappa & $\underset{(0.1186)}{\underline{0.7851}}$ & $\underset{(0.1186)}{0.7362}$ & $\underset{(0.0355)}{\mathbf{0.9451}}$ & $\underset{(0.0674)}{0.6646}$ & $\underset{(0.0645)}{0.6551}$ & $\underset{(0.0850)}{0.4264}$ & $\underset{(0.0860)}{0.4290}$ \\ 
\cline{2-10}
\specialrule{0em}{1pt}{1pt}
& \multirow{7}{*}{RF} & ACC & $\underset{(0.0763)}{\underline{0.8600}}$ & $\underset{(0.0763)}{0.8333}$ & $\underset{(0.0450)}{\mathbf{0.9238}}$ & $\underset{(0.0547)}{0.8250}$ & $\underset{(0.0569)}{0.8213}$ & $\underset{(0.0763)}{0.5043}$ & $\underset{(0.0867)}{0.5088}$ \\ 
&  & recall & $\underset{(0.0763)}{\underline{0.8600}}$ & $\underset{(0.0763)}{0.8333}$ & $\underset{(0.0450)}{\mathbf{0.9238}}$ & $\underset{(0.0547)}{0.8250}$ & $\underset{(0.0569)}{0.8213}$ & $\underset{(0.0763)}{0.5043}$ & $\underset{(0.0867)}{0.5088}$ \\ 
&  & $F_1$ & $\underset{(0.0886)}{\underline{0.8244}}$ & $\underset{(0.0886)}{0.7924}$ & $\underset{(0.0571)}{\mathbf{0.9009}}$ & $\underset{(0.0652)}{0.7808}$ & $\underset{(0.0683)}{0.7751}$ & $\underset{(0.0769)}{0.4369}$ & $\underset{(0.0853)}{0.4428}$ \\ 
&  & Kappa & $\underset{(0.0783)}{\underline{0.8564}}$ & $\underset{(0.0783)}{0.8290}$ & $\underset{(0.0462)}{\mathbf{0.9218}}$ & $\underset{(0.0561)}{0.8205}$ & $\underset{(0.0584)}{0.8167}$ & $\underset{(0.0783)}{0.4915}$ & $\underset{(0.0889)}{0.4962}$ \\ 
\cline{2-10}
\specialrule{0em}{1pt}{1pt}
\multirow{7}{*}{ORL} & \multirow{7}{*}{DT} & ACC & $\underset{(0.0751)}{\mathbf{0.5815}}$ & $\underset{(0.0751)}{\underline{0.5463}}$ & $\underset{(0.0691)}{0.4678}$ & $\underset{(0.0640)}{0.4360}$ & $\underset{(0.0800)}{0.4420}$ & $\underset{(0.0863)}{0.3513}$ & $\underset{(0.0818)}{0.3525}$ \\ 
&  & recall & $\underset{(0.0751)}{\mathbf{0.5815}}$ & $\underset{(0.0751)}{\underline{0.5463}}$ & $\underset{(0.0691)}{0.4678}$ & $\underset{(0.0640)}{0.4360}$ & $\underset{(0.0800)}{0.4420}$ & $\underset{(0.0863)}{0.3513}$ & $\underset{(0.0818)}{0.3525}$ \\ 
&  & $F_1$ & $\underset{(0.0744)}{\mathbf{0.5159}}$ & $\underset{(0.0744)}{\underline{0.4771}}$ & $\underset{(0.0706)}{0.4022}$ & $\underset{(0.0630)}{0.3695}$ & $\underset{(0.0784)}{0.3760}$ & $\underset{(0.0834)}{0.2935}$ & $\underset{(0.0772)}{0.2927}$ \\ 
&  & Kappa & $\underset{(0.077)}{\mathbf{0.5708}}$ & $\underset{(0.0770)}{\underline{0.5346}}$ & $\underset{(0.0709)}{0.4541}$ & $\underset{(0.0656)}{0.4215}$ & $\underset{(0.0821)}{0.4277}$ & $\underset{(0.0885)}{0.3346}$ & $\underset{(0.0839)}{0.3359}$ \\ 
\cline{2-10}
\specialrule{0em}{1pt}{1pt}
& \multirow{7}{*}{lightGBM} & ACC & $\underset{(0.0773)}{\underline{0.7848}}$ & $\underset{(0.0773)}{0.7458}$ & $\underset{(0.0574)}{\mathbf{0.7968}}$ & $\underset{(0.0721)}{0.6133}$ & $\underset{(0.0708)}{0.6163}$ & $\underset{(0.0851)}{0.4508}$ & $\underset{(0.0821)}{0.4483}$ \\ 
&  & recall & $\underset{(0.0773)}{\underline{0.7848}}$ & $\underset{(0.0773)}{0.7458}$ & $\underset{(0.0574)}{\mathbf{0.7968}}$ & $\underset{(0.0721)}{0.6133}$ & $\underset{(0.0708)}{0.6163}$ & $\underset{(0.0851)}{0.4508}$ & $\underset{(0.0821)}{0.4483}$ \\ 
&  & $F_1$ & $\underset{(0.0867)}{\underline{0.7363}}$ & $\underset{(0.0867)}{0.6896}$ & $\underset{(0.0675)}{\mathbf{0.7470}}$ & $\underset{(0.0801)}{0.5434}$ & $\underset{(0.0764)}{0.5479}$ & $\underset{(0.0831)}{0.3836}$ & $\underset{(0.0814)}{0.3814}$ \\ 
&  & Kappa & $\underset{(0.0793)}{\underline{0.7792}}$ & $\underset{(0.0793)}{0.7392}$ & $\underset{(0.0588)}{\mathbf{0.7915}}$ & $\underset{(0.0739)}{0.6033}$ & $\underset{(0.0726)}{0.6064}$ & $\underset{(0.0873)}{0.4367}$ & $\underset{(0.0842)}{0.4341}$ \\ 
\cline{2-10}
\specialrule{0em}{1pt}{1pt}
& \multirow{7}{*}{MLP} & ACC & $\underset{(0.0736)}{\underline{0.8420}}$ & $\underset{(0.0736)}{0.8055}$ & $\underset{(0.0484)}{\mathbf{0.8638}}$ & $\underset{(0.0674)}{0.7000}$ & $\underset{(0.0668)}{0.6973}$ & $\underset{(0.0908)}{0.4615}$ & $\underset{(0.0881)}{0.4670}$ \\ 
&  & recall & $\underset{(0.0736)}{\underline{0.8420}}$ & $\underset{(0.0736)}{0.8055}$ & $\underset{(0.0484)}{\mathbf{0.8638}}$ & $\underset{(0.0674)}{0.7000}$ & $\underset{(0.0668)}{0.6973}$ & $\underset{(0.0908)}{0.4615}$ & $\underset{(0.0881)}{0.4670}$ \\ 
&  & $F_1$ & $\underset{(0.0851)}{\underline{0.8038}}$ & $\underset{(0.0851)}{0.7597}$ & $\underset{(0.0583)}{\mathbf{0.8286}}$ & $\underset{(0.0727)}{0.6398}$ & $\underset{(0.0723)}{0.6372}$ & $\underset{(0.0882)}{0.3928}$ & $\underset{(0.0847)}{0.3992}$ \\ 
&  & Kappa & $\underset{(0.0755)}{\underline{0.8379}}$ & $\underset{(0.0755)}{0.8005}$ & $\underset{(0.0496)}{\mathbf{0.8603}}$ & $\underset{(0.0691)}{0.6923}$ & $\underset{(0.0686)}{0.6895}$ & $\underset{(0.0931)}{0.4477}$ & $\underset{(0.0904)}{0.4533}$ \\ 
\bottomrule
\label{tab:appendix_A1}
\end{longtable}
\end{center}

\section*{Appendix C} \label{appendix_2}
\setcounter{table}{0}
\setcounter{figure}{0}
\renewcommand{\thetable}{C\arabic{table}}
\renewcommand{\thefigure}{C\arabic{figure}}
	
	\begin{figure}[H]
	\centering
	\includegraphics[width=0.87\linewidth]{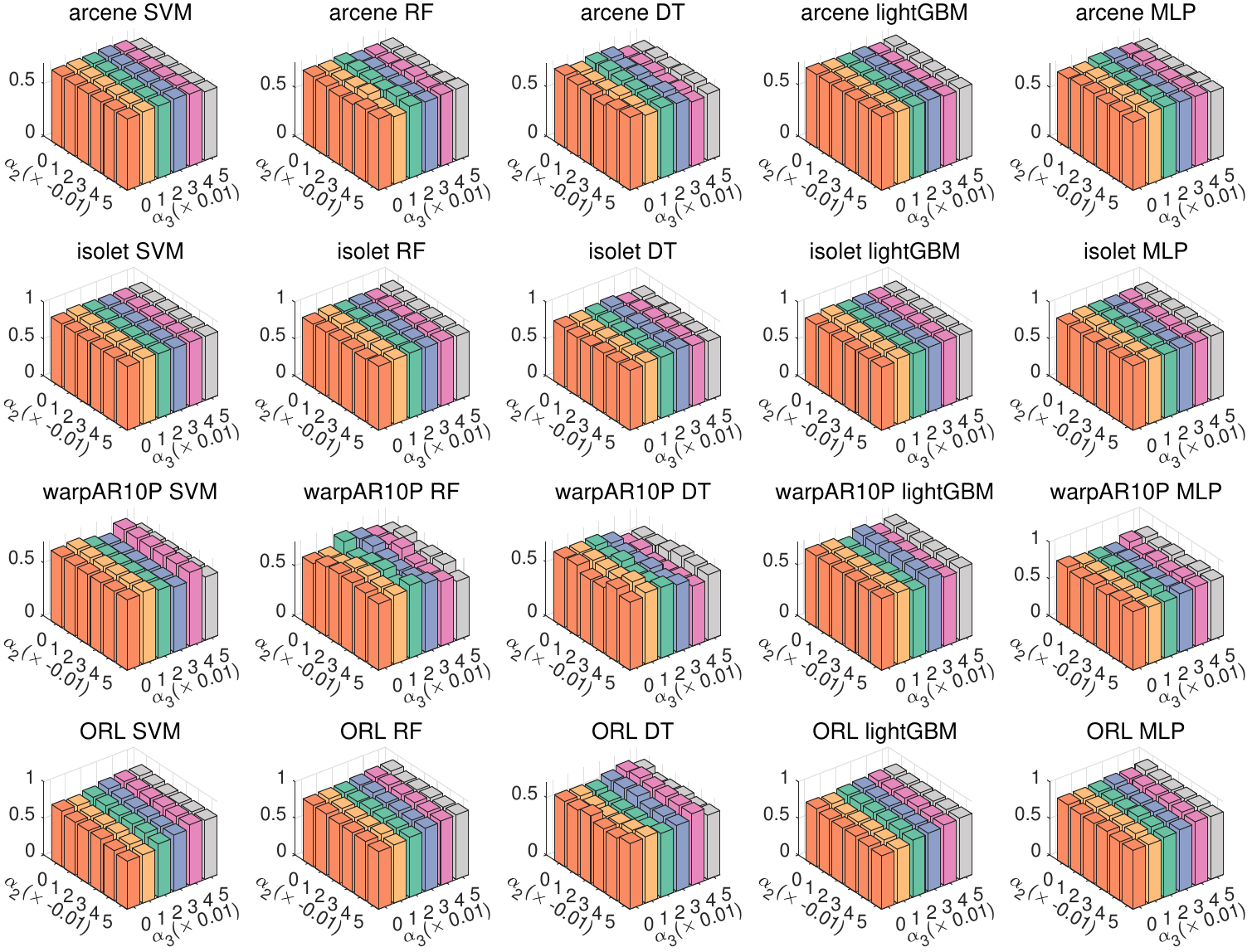}
	\caption{The accuracy of different learning models using the selected predictors under different parameter settings of TNVS.}
	\label{fig:parameter_adj_pred}
\end{figure}


	\par\noindent 
	
\end{document}